\documentclass[aps,pra,twocolumn,superscriptaddress,amsfont,amssymb,amsmath,nofootinbib,showpacs,balancelastpage,english,10pt]{revtex4-1} 
\usepackage{babel}
\usepackage{graphicx}
\usepackage{dcolumn}% Align table columns on decimal point
\usepackage{bm}% bold math
\usepackage{color}

\begin{document}

\title{Weak value amplification and beyond the standard quantum limit in position measurements}

\author{Atsushi Nishizawa}
\email{anishi@caltech.edu}
\affiliation{Theoretical Astrophysics 350-17, California Institute of Technology, Pasadena, California 91125, USA}

\begin{abstract}%%%%%%%%%%%%%%%%%%%%%%%%%%%%%%%%%%%%%
In a weak measurement with post-selection, a measurement value, called the weak value, can be amplified beyond the eigenvalues of the observable. However, there are some controversies whether the weak value amplification is practically useful or not in increasing sensitivity of the measurement in which fundamental quantum noise dominates. In this paper, we investigate the sensitivity limit of an optical interferometer by properly taking account quantum shot noise and radiation pressure noise. To do so, we formulate the weak value amplification in the Heisenberg picture, which enables us to intuitively understand what happens when the measurement outcome is post-selected and the weak value is amplified. As a result, we found that the sensitivity limit is given by the standard quantum limit that is the same as in a standard interferometry. We also discuss a way to circumvent the standard quantum limit.
\end{abstract}

\date{\today}

\maketitle

%%%%%%%%%%%%%%%%%%%%%%%%%%%%%%%%%%%%%
\section{Introduction}

The idea of weak-value amplification (WVA) was originally introduced by Aharonov, Albert, and Vaidman (AAV) in 1988 \cite{Aharonov:1988xu} (see \cite{Aharonov:2007LNP} for a review). When a system is weakly measured by a measuring device, the measurement results can be much larger than the eigenvalues of the observable by appropriately selecting initial and final states of the system. This theoretical prediction has been demonstrated in various pioneering experiments, e.g. the rotation of photon polarization \cite{Ritchie:1991,Pryde:2004zw}, quantum box problem \cite{Resch:2004PhLA}, the arrival time of a single photon \cite{Wang:2006PRA}, the spin Hall effect of light \cite{Hosten:2008}, optical beam deflection \cite{Dixon:2009PRL,Starling:2009PRA}, and optical phase \cite{Starling:2010aPRA}.
For further recent theoretical and experimental developments on the WVA, see review papers \cite{Kofman:2012rev,Dressel:2014rev,Shikano:2011}.

The interesting nature of the WVA results from the definition of the weak value:
\begin{equation}
A_w \equiv \frac{\langle \psi_f | A |\psi_i \rangle}{\langle \psi_f | \psi_i \rangle} \;, 
\label{eq50}
\end{equation}
where $A$ is an observable associated with the system to be measured, $|\psi_{i}\rangle$ and $|\psi_{f}\rangle$ are the initial and final states of the system. The weak value is interpreted as the observable evaluated at intermediate times between the pre- and post-selections. From this definition, we see that if the pre- and post-selected states are nearly orthogonal, it seems that the weak value becomes arbitrarily large. In fact, the experiments mentioned above have shown that the WVA significantly improves signal-to-noise ratio (SNR) in the situation where technical noise (e.g. alignment noise) dominates and outperforms a standard interferometry. The theoretical study also confirms this advantage of the WVA, e.g.~\cite{Brunner:2010,Jordan:2014PRX}, though it has been pointed out that the applicability of the amplification strongly depends on the property of technical noise \cite{Knee2014PRX}. On the other hand, when the weak value becomes large, we must take into account the nonlinear effects of the von-Neumann measurement \cite{Lorenzo:2008,Zhu:2011PRA,Nakamura:2012,Koike:2011}. As the result, the amplification factor has a maximum value and vanishes when the pre- and post-selected states are exactly orthogonal. Even in the weak measurement regime, the nearly orthogonal post-selection severely reduces output statistics, which consequently compensates the improvement of sensitivity due to the amplification of the signal. On the other hand, some authors \cite{Parks:2011,Susa:2012} have claimed that the sensitivity is infinitely improved by optimizing the wave functions of a system and a probe. However, if quantum shot noise in the detection process is taken into consideration, their conclusion would change. Therefore, the practical usefulness of the WVA in increasing sensitivity has been often controversial.

In an optical interferometry, it was pointed out in \cite{Starling:2009PRA,Nishizawa:2012weak} that if photon shot noise dominates, there is no advantage to use the WVA for improving the fundamental limit of parameter estimation precision. From an informational approach with quantum Fisher information \cite{Tanaka:2013PRA,Ferrie:2014PRL,Combes:2014PRA}, it is also obtained the same conclusion that the WVA is suboptimal in enhancing the estimation precision and does not perform better than the standard statistical techniques (for the controversy on this conclusion, see also \cite{Vaidman:2014comment,Kedem:2014comment,Ferrie:2014comment}). Indeed, these conclusions have been demonstrated in an optical experiment \cite{Viza:2014}. However, in an optical interferometry, not only shot noise but also radiation pressure noise contribute to fundamental quantum noise and leads to a kind of sensitivity limit, the so-called standard quantum limit (SQL) \cite{Braginsky:1975SvPhU,Caves:1980rv}. It is not clear how radiation pressure noise affects the sensitivity limit of the measurement and alters the SQL from that in the standard interferometry when the WVA is implemented. These questions are worth investigating because there are many systems, in which the radiation pressure and the SQL play an important role, such as optomechanical systems \cite{Aspelmeyer:2014rev,Poot:2012rev} and gravitational-wave detection \cite{Kimble:2000gu,Danilishin:2012rev}. 

Conventionally, the weak measurement with post-selection and the WVA are formulated in the Schr\"odinger picture. However, this is somewhat less intuitive, compared with the Heisenberg formulation, and often leads to confusion. Also it is more difficult to deal with moving mirrors and the radiation pressure noise in the Schr\"odinger picture. Instead, here we formulate the WVA in the Heisenberg picture and deal with the quantum noises in a fully quantum-mechanical way, including the radiation pressure noise. Then we derive the SQL and discuss a method to circumvent beyond the SQL. 

This paper is organized as follows. In Sec.~\ref{sec2}, we briefly review position measurements in an optical interferometer and the WVA in the Schr\"odinger picture. In Sec.~\ref{sec3}, we formulate the WVA  in the Heisenberg picture, appropriately taking into account quantum shot noise and radiation pressure noise. The SQL is derived in Sec.~\ref{sec4} and the method to overcome the SQL is presented in Sec.~\ref{sec5}. Finally, Sec.~\ref{sec6} is devoted to a summary. In this paper, we use the unit $c=1$.

%%%%%%%%%%%%%%%%%%%%%%%%%%%%%%%%%%%%%
\section{Weak value amplificaiton of a signal in the Schr\"odinger picture}
\label{sec2} 

\begin{figure}[h]
\begin{center}
\includegraphics[width=7cm]{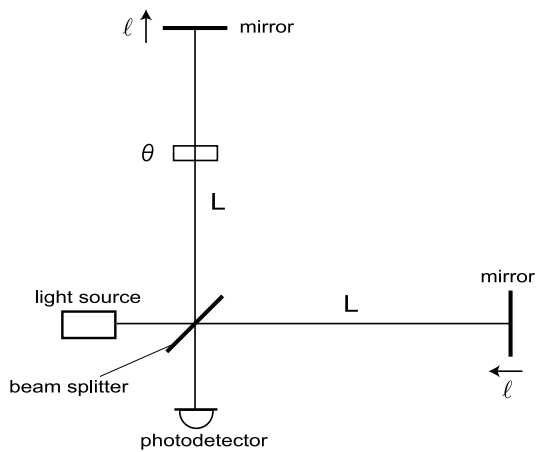}
\caption{Michelson interferometer.}
\label{fig4b}
\end{center}
\end{figure}

We briefly review the Schr\"odinger formulation of WVA, which is the original formulation by Aharonov {\it{et al.}}~\cite{Aharonov:1988xu} and conventionally used so far. In this paper, we concentrate on a small displacement measurement of mirrors in an optical interferometer, particularly focusing on a Michelson interferometer shown in Fig.~\ref{fig4b}. This does not loose generality because a Michelson interferometer is geometrically equivalent to a Mach-Zehnder interferometer and a Sagnac interferometer and gives the same sensitivity limit of a position measurement. 

We start with a single photon case for an illustrative purpose. (We will generalize the result to a macroscopic beam later.) A photon is injected from the left side, which we call the bright port because we will extend later to the case of a laser light pulse. The photon that enters the interferometer takes two paths with the same unperturbed distance $L$ after being divided by the beam splitter. One of the photons (beams) in the y-arm is phase-shifted by $\theta$ (pre-selection), reflected at the mirror with sensing small displacements $\ell$ of the mirror (weak interaction), and recombined at the beam splitter with another photon that sensed small mirror displacement $-\ell$ in the x-arm (post-selection). Here we are interested in small differential displacement of the mirrors since we focus on the detection of external disturbances on the mirrors. Finally, the photon is detected by the photodetector at the output (dark) port.

In this optical setup, we regard the beam's which-path information (y-arm or x-arm) as the system to be measured via a weak measurement. The initial pre-selected state of the system is denoted by 
\begin{equation}
| \psi_i \rangle = \frac{1}{\sqrt{2}} \left( e^{i \theta /2} | y \,\rangle + e^{-i \theta /2} |x \,\rangle \right) \;. \nonumber
\end{equation}
The initial phase offset $\theta$ is symmetrized merely for simplicity of calculation. An observable of an measuring device, or a pointer variable, is photon's momentum (frequency), which measures the phase shift induced by the mirror displacements in the Michelson interferometer. The initial state of the pointer is
\begin{equation}
| \Phi \rangle = \int dp\, \Phi(p) |p \rangle \;. \nonumber
\end{equation}
Since we measure the small displacement of the mirrors at the asymmetric output port of this optical configuration, the observable is $2 \ell A$, where the operator $A \equiv | y \rangle \langle y |-| x \rangle \langle x | $ carries information about which arm a photon passes and has the eigenvalues $\pm 1$, depending on the photon path. 

The Hamiltonian of the interaction, which is switched on at time $t_0$, is written as  
\begin{equation}
H = g \delta(t-t_0) \, A \otimes p \;. 
\label{eq2}
\end{equation}
Here we defined $g \equiv -2 \ell$. This interaction Hamiltonian is interpreted as a generalization of von Neumann interaction like $\ell \otimes p$ to the Michelson-type interferometer. After the interaction given in Eq.~(\ref{eq2}) and the post-selection by the final state of the system, $|\psi_f \rangle =( | y \rangle - | x \rangle )/\sqrt{2}$, the final state of the device can be exactly evaluated including nonlinear terms in the coupling \cite{Nishizawa:2012weak},
\begin{align}
|\Phi^{'} \rangle &= \langle \psi_f | e^{-i g A p} |\psi_i \rangle | \Phi \rangle \nonumber \\
&= \int dp \,\Phi (p) |p \rangle \langle \psi_f |\psi_i \rangle (\cos gp -i A_w \sin gp) \;, 
\label{eq4}
\end{align}

The expectation value of the $n$-th power of $p$ for the final state of the photon in Eq.~(\ref{eq4}) is given by \cite{Nishizawa:2012weak} 
\begin{equation}
\langle p^n \rangle^{'} = \frac{\langle p^n \rangle + (|A_w|^2-1) \langle p^n \sin^2 gp \rangle + {\rm{Im}} A_w \langle p^n \sin 2gp \rangle}{1+ (|A_w|^2-1) \langle \sin^2 gp \rangle + {\rm{Im}} A_w \langle \sin 2gp \rangle} \;. 
\label{eq5}
\end{equation}
The bracket $\langle \cdots \rangle$ and $\langle \cdots \rangle^{'}$ denote averaging over the initial and final state of the measuring device, respectively. In our case, the weak value defined in Eq.~(\ref{eq50}) is
\begin{equation}
A_w = -i \cot \frac{\theta}{2} \;.
\label{eq6}
\end{equation}

The series of power in Eq. (\ref{eq5}) contains all information about the system. If we measure the shift of the pointer variable $\langle p \rangle^{'}$, the variance ${\rm{Var}} [p]^{'} = \langle p^2 \rangle^{'} - (\langle p \rangle^{'})^2$ is regarded as frequency noise, because it results from a photon spectral distribution. However, in an optical experiment, shot noise coming from the fluctuation of the photon number also contributes. As shown in \cite{Nishizawa:2012weak}, taking into account the photon number fluctuations, it turned out to be that the shot noise is given by $\langle p^2 \rangle^{'}$ and is always larger than the frequency noise. Therefore, it is essential to consider the shot noise when we discuss sensitivity limit in an optical interferometer.

Hereafter we denote $p$ by $\hbar \omega$ as we are considering an optical experiment. Given the initial momentum distribution of a photon is non-zero-mean Gaussian (For multiple photons, a pulsed laser whose central frequency is mode-locked to $\omega_0$), 
\begin{equation}
\Phi  (\omega) = \left( \frac{1}{2\pi \sigma_{\omega}^2} \right)^{1/4} \exp \left[ -\frac{(\omega-\omega_0)^2}{4 \sigma_{\omega}^2} \right] \;, \nonumber
\end{equation}
and substituting Eq.~(\ref{eq6}) for Eq.~(\ref{eq5}), we obtain the following expressions for the first and second powers of $\omega-\omega_0$ \footnote{The sign of $\phi$ is different from that in \cite{Nishizawa:2012weak}. This is just the matter of a different sign for mirror displacement.}:
\begin{align}
g \langle \omega - \omega_0 \rangle^{'} &= \frac{s\, e^{-s} \sin (\theta+\phi)}{1-e^{-s} \cos (\theta +\phi )} \;, 
\label{eq11a} \\
g^2 \langle (\omega- \omega_0)^2 \rangle^{'} &= \frac{s}{2} \left[ 1+ \frac{2s\,e^{-s}\,\cos (\theta+\phi)}{1-e^{-s} \cos (\theta +\phi)} \right] 
\label{eq12a} \;.
\end{align} 
where
\begin{equation}
s \equiv 2 g^2 \sigma_{\omega}^2 = 8 \sigma_{\omega}^2 \ell^2 \;, \quad \quad \phi \equiv -2 g \omega_0 = 4 \omega_0 \ell \;. 
\label{eq45}
\end{equation}
The parameter $s$ characterizes the measurement strength, since large $g$ means strong coupling of the interaction and large $\sigma_{\omega}$ means the narrow distribution of photons in the time domain. Then the weak measurement is defined by the limit, $s \rightarrow 0$, which corresponds to two physical situations: (i) the measurement coupling, in this case mirror displacement, is small, (ii) probe's wave function is narrow in momentum (frequency) space or broad in spatial (time) domain. 

At the limit of weak measurement, if only linear terms in $s$ are kept, the above equations, (\ref{eq11a}) and (\ref{eq12a}), are reduced to
\begin{align} 
g \langle \omega - \omega_0 \rangle^{'} &\approx  s \cot (\theta + \phi)  \;, \nonumber \\
g^2 \langle (\omega- \omega_0)^2 \rangle^{'} &\approx \frac{s}{2} \;.
\end{align}
From these expressions, one finds that the frequency shift is proportional to $\cot (\theta + \phi)$ and could be amplified for small $\phi$ by taking small $\theta$. Note, however, that these expressions are valid only in the linear regime of $s$ and that sensitivity is not infinitely amplified if nonlinear orders are included.

The above results can be easily extended to multiple-photons. From Eq.~(\ref{eq5}), the probability distribution for a single photon at the output is 
\begin{equation}
\rho(\omega) \equiv \frac{|\langle \omega | \Phi^{'} \rangle|^2}{\langle \Phi^{'} | \Phi^{'} \rangle} \;. 
\end{equation}
For $N$ output photons, the photon number distribution is simply given by
\begin{equation}
\bar{n}(\omega) = N \rho (\omega)\;. 
\end{equation}
However, the extension of the wave function from one photon to multiple photons is somewhat conceptually strange. In the Heisenberg picture we discuss in the next section, the photon number distribution at the output is more naturally introduced.
Also in this discussion, radiation pressure noise is not included. To fully take into account the radiation pressure noise, we need to move to the Heisenberg picture.

%Suppose that a coherent light with multiple frequency modes is injected from the bright port and the dark port is in a coherent vacuum state, though we will extend to a squeezed vacuum later. Using the displacement operator ${\cal{D}}$, the initial state is given by
%\begin{align}
%|\psi \rangle &= \prod_i | \alpha_i \rangle_d \otimes |0\rangle_a \nonumber \\
%&= {\cal{D}} |0\rangle_d \otimes |0\rangle_a  \;, 
%\label{eq21}
%\end{align}
%where the displacement operator for the $d$-field is defined by
%\begin{equation}
%{\cal{D}} \equiv \exp \left[ \int \frac{d\omega}{2\pi} \alpha(\omega) D^{\dag}%(\omega) -\alpha^{\ast} (\omega) D(\omega) \right] \;.
%\end{equation}
%and $D(\omega)$ and $D^{\dag}(\omega)$ are the annihilation and creation operators of the $d$-field.

%%%%%%%%%%%%%%%%%%%%%%%%%%%%%%%%%%%%%
\section{Weak value amplification of a signal in the Heisenberg picture}
\label{sec3} 

When we deal with a radiation pressure force, it is convenient to work in the Heisenberg picture. To do so, we use the single-photon formalism for a quantum electromagnetic field in an interferometer, in contrast to the two-photon formalism in Kimble et al. \cite{Kimble:2000gu}, because a pulsed laser has a broad spectrum around a central frequency. In the Heisenberg picture, an electric field is written as 
\begin{align}
E(t) &= E^{(+)}(t) + E^{(-)} (t)\;, \label{eq22} \\
E^{(+)} (t)
&= \int_0^{\infty} \sqrt{\frac{2 \pi \hbar \omega}{{\cal{A}}\,c}}\; a_\omega\,
e^{-i\omega t} \frac{d\omega }{ 2\pi}\;, \label{eq24}  \\
E^{(-)} &= [E^{(+)}]^{\dag} \;. \label{eq25}
\end{align}
Here ${\cal{A}}$ is the effective scattering cross-section of the beam, $a_{\omega}$ is the annihilation operator of a positive frequency mode $\omega$, which satisfies the following commutation relations
\begin{equation}
[a_\omega, a_{\omega^{\prime}}] = 0\;, \quad \quad
[a_\omega, a_{\omega'}^{\dag} ] = 2\pi \delta(\omega-\omega^{\prime})\;. \nonumber
\end{equation}

Let us denote the input (incoming) and output (outgoing) fields at the dark port by $a$ and $b$, respectively. Suppose that a laser pulse is injected from the bright port and the field is denoted by $D=\alpha + d$ with a classical part $\alpha$ and a quantum fluctuating part $d$. For brevity, we omit the subscript $\omega$ for the field hereafter. The input-output relation for the fields is derived in Appendix \ref{app:io-relation} and is given up to the linear order in vacuum fluctuations by 
\begin{align}
b&\approx i\, e^{2i \omega \tau} \left\{ \alpha \sin \omega \xi + \alpha \omega \xi_{\rm{r}} \cos \omega \xi  + d\sin \omega \xi -i a \cos \omega \xi \right\} \;,
\end{align}
where $\xi \equiv 2\ell + \tau_{\theta}$
and 
\begin{equation}
\xi_{\rm{r}} = \frac{2 T}{m} \int_{0}^{\infty} \frac{d\omega}{2\pi} \hbar \omega (\alpha a^{\dag} + \alpha^{*} a) \;. \label{eq46}
\end{equation}
The definition of $\tau_{\theta}$ is just for convenience of notation and is determined so that $\theta/2 \equiv \omega \, \tau_{\theta}(\omega)$. The time shift $\xi_{\rm{r}}$ comes from radiation pressure force exerted on a mirror. In the expression, $m$ is the mirror mass and $T$ is the interval of measurements. The $T$
dependence shows up in $\xi_{\rm{r}}$ merely because the mirror moves by the momentum change due to the disturbance of the previous measurement. In the derivation, we assumed $|\omega \xi_{\rm{r}}| \ll 1$ and neglected the higher order terms in the vacuum fields and $\omega \xi_{\rm{r}}$. 

The photon number at the output is 
\begin{align}
n &= b^{\dag}b \nonumber \\
&\approx  |\alpha|^2 \sin^2 \omega \xi + 2 |\alpha|^2  \omega \xi_{\rm{r}} \cos \omega \xi \sin \omega \xi \nonumber \\
&+( \alpha^{*} d + \alpha d^{\dag}) \sin^2 \omega \xi + i ( \alpha a^{\dag} - \alpha^{*} a ) \sin \omega \xi \cos \omega \xi \;.
\label{eq13}
\end{align}
Evaluating this in a coherent vacuum state $| 0 \rangle \equiv |0\rangle_d |0\rangle_a$, we have the average number of photons at the output
\begin{align}
\bar{n}(\omega) &= \langle 0 | n | 0 \rangle \nonumber \\
&= \bar{n}_0 (\omega) \sin^2 \omega \xi \;, \label{eq23} 
\end{align}
where we defined the average photon number distribution at the input by $\bar{n}_0 (\omega) \equiv |\alpha(\omega)|^2$.

In what follows, we consider the Gaussian wave packet as is the same in the case of Schr\"odinger picture. However, it is not a wave function of a photon but here the photon number distribution in the Heisenberg picture 
\begin{equation}
\alpha (\omega) = A \exp \left[- \frac{(\omega - \omega_0)^2}{4\sigma_{\omega}^2} \right] \;, 
\end{equation}
for the frequency $\omega >0$. Then the spread should be $\sigma_{\omega} \lesssim \omega_0$. At $\omega \approx 0$, the distribution has an abrupt cutoff. This is somewhat artificial, but in most cases the cutoff is at the tail of the Gaussian distribution and does not much affect the observable signature at the output only if $\sigma_{\omega} \lesssim \omega_0$. Defining a sideband, $\omega=\omega_0+\Omega$, and the normalized quantities $\tilde{\Omega} \equiv \Omega/\omega_0$ and $\tilde{\sigma}_{\omega} \equiv \sigma_{\omega}/\omega_0$, the distribution is written as
\begin{equation}
\alpha(\tilde{\Omega}) = A \exp \left[- \frac{\tilde{\Omega}^2}{4\tilde{\sigma}_{\omega}^2} \right] \;,
\end{equation}
for $\tilde{\Omega} \geq - 1$ and $\tilde{\sigma}_{\omega} \leq 1$. Then the average input power of a pulse is
\begin{align}
P_0 &= \int_{0}^{\infty} \frac{d\omega}{2\pi} \hbar \omega \, \bar{n}_0 (\omega) \nonumber \\
&\approx \hbar \omega_0^2 |A|^2 \int_{-\infty}^{\infty} \frac{d\tilde{\Omega}}{2\pi} (1+\tilde{\Omega}) e^{-\tilde{\Omega}^2/2\tilde{\sigma}_{\omega}^2} \nonumber \\
&= \frac{1}{\sqrt{2\pi}} \hbar \omega_0 \sigma_{\omega} |A|^2 \;. 
\label{eq26}
\end{align}
At the second line, although the range of $\tilde{\Omega}$ is limited to be $\tilde{\Omega} \geq -1$ for the probe light, we extended the integral range to $-\infty \leq \tilde{\Omega} \leq \infty$ because the contribution from the frequencies $\tilde{\Omega} < -1$ is exponentially suppressed owing to the Gaussian tail.

Using the definition of the phase shift due to mirror displacement in Eq.~(\ref{eq45}) and writing the phase factor in the signal as
\begin{equation}
\omega \xi = (1+\tilde{\Omega})\frac{\phi}{2} + \frac{\theta}{2} \;,
\end{equation}
we have the averaged photon-number distribution at the output in the form
\begin{equation}
\bar{n}(\tilde{\Omega}) = \frac{\sqrt{2\pi}}{\omega_0 \tilde{\sigma}_{\omega}} N_0  e^{-\tilde{\Omega}^2/2 \tilde{\sigma}_{\omega}^2} \sin^2 \left[ \frac{\theta}{2} +(1+\tilde{\Omega} ) \frac{\phi}{2} \right] \;.
\label{eq37}
\end{equation}
where $N_0 \equiv P_0/\hbar \omega_0$ is the effective total number of photons at the input. Note that the actual number of photon is different from $N_0$ since frequency is not monochromatic. All photon number at the output is obtained by integrating the photon distribution over all frequencies, $-1 \leq \tilde{\Omega} \leq \infty$. As well as the above integral in Eq.~(\ref{eq26}), we extend the integral range to $-\infty \leq \tilde{\Omega} \leq \infty$ and use the mathematical formulas for the integrals, Eqs.~(\ref{eqc1}) and (\ref{eqc2}), we have the total number of photons at the output
\begin{align}
N &\approx \omega_0 \int_{-\infty}^{\infty} \frac{d\tilde{\Omega}}{2\pi} \, \bar{n}(\tilde{\Omega}) \nonumber \\
&= \frac{N_0}{2} \left[ 1-e^{- \tilde{\sigma}_{\omega}^2 \phi^2/2} \cos (\theta +\phi) \right] \;.
\label{eq15}  
\end{align}

The quantity $\tilde{\sigma}_{\omega}^2 \phi^2/2$ appearing in Eq.~(\ref{eq15}) coincides with the measurement strength parameter $s$ in Eq.~(\ref{eq45}). When the measurement is weak ($s \ll 1$) and the post-selection is nearly orthogonal ($|\theta| \ll 1$), the output photon number is significantly suppressed from the input photon number. On the other hand, for $|\theta| \ll 1$, the weak value defined in Eq.~(\ref{eq6}) is largely amplified. Therefore, we observe that there is a tradeoff between the magnitude of the signal and its statistics.

Once we have the photon number distribution at the output, it is straightforward to compute the frequency shift and its variance. The frequency shift is
\begin{align}
\langle \tilde{\Omega} \rangle &\approx \omega_0 \int_{-\infty}^{\infty} \frac{d\tilde{\Omega}}{2\pi} \, \tilde{\Omega}\, \frac{\bar{n}(\tilde{\Omega})}{N}  \nonumber \\
&= \frac{2s}{\phi} \frac{e^{-s} \sin (\theta+\phi)}{1-e^{-s}\cos (\theta+\phi)} \;.
\label{eq20}
\end{align}
At the first line, we again approximately extended the integral range to from $-\infty$. Since $\langle \tilde{\Omega} \rangle \phi/2 = g  \langle \Omega \rangle$, this coincides with the result in the Schr\"odinger picture in Eq.~(\ref{eq11a}). As well, the expectation value of squared frequency shift is
\begin{align}
\langle \tilde{\Omega}^2 \rangle &\approx \omega_0 \int_{-\infty}^{\infty} \frac{d\tilde{\Omega}}{2\pi} \, \tilde{\Omega}^2\, \frac{\bar{n}(\tilde{\Omega})}{N} \nonumber \\
&= \frac{2s}{\phi^2} \left[ 1+\frac{2s\, e^{-s} \cos (\theta+\phi)}{1-e^{-s} \cos (\theta+\phi)} \right] \;,
\label{eq38} 
\end{align}
Again this coincides with Eq.~(\ref{eq12a}) obtained in the Schr\"odinger picture.

\begin{figure}[t]
\begin{center}
\includegraphics[width=7.5cm]{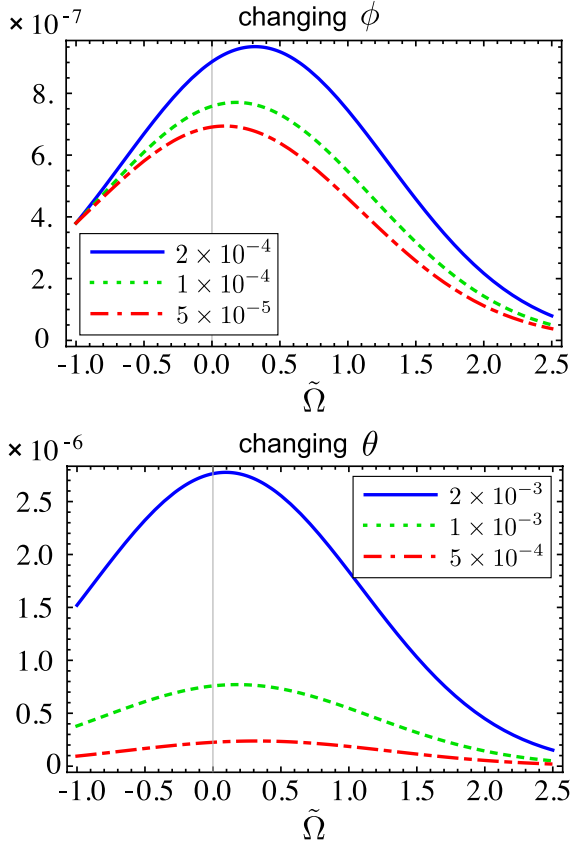}
\caption{Photon number distribution at the output $\bar{n}(\tilde{\Omega})$ in the unit of $N_0/\omega_0$. In each panel, one parameter is varied with fixing fiducial parameter to $\tilde{\sigma}_{\omega}=1$, $\phi=10^{-4}\,{\rm{rad}}$, and $\theta=10^{-3}\,{\rm{rad}}$. In the upper panel, $\phi=2\times 10^{-4}$ (blue, solid), $10^{-4}$ (green, dotted), and $5\times 10^{-5}\,{\rm{rad}}$ (red, dotted-dashed). In the lower panel, $\theta=2\times 10^{-3}$ (blue, solid), $10^{-3}$ (green, dotted), and $5\times 10^{-4}\,{\rm{rad}}$ (red, dotted-dashed). }
\label{fig1a}
\end{center}
\end{figure}

\begin{figure}[t]
\begin{center}
\includegraphics[width=7.5cm]{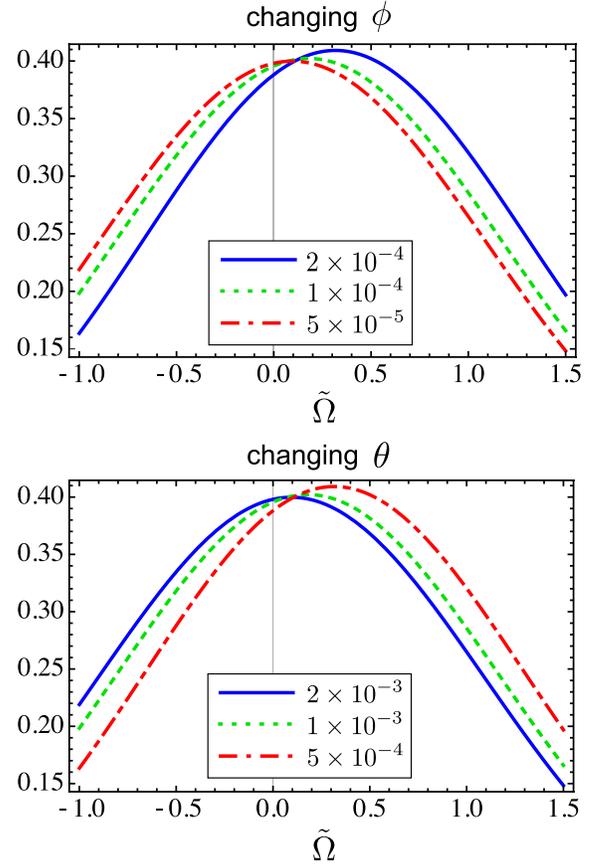}
\caption{Probability distribution at the output $\omega_0 \bar{n}(\tilde{\Omega})/(2\pi N)$. The parameters for each curve in each panel are the same as Fig.~\ref{fig1a}, but they are normalized by the total photon number at the output.}
\label{fig1b}
\end{center}
\end{figure}

Figure \ref{fig1a} shows the photon number distribution at the output $\bar{n}(\tilde{\Omega})$ in the unit of $N_0/\omega_0$, varying $\phi$ and $\theta$, respectively. As $\phi$ increases, the distribution is amplified and more shifted. On the other hand, as $\theta$ decreases (the weak value is more amplified), the distribution is more shifted but significantly suppressed. These behaviors indicate the existence of the trade-off relation between the amplification of the weak value and the amount of output statistics. In the Heisenberg formulation, the role of the post-selection is much clearer than the Schr\"odinger formulation. In Fig.~\ref{fig1b}, the photon number distribution is normalized and then is interpreted as probability distribution. If one see the probability distribution, the peak frequency is shifted without suppression of the amplitude. This feature often causes confusion with the amplification of the SNR as in the Schr\"odinger picture. 

In this section, we considered the average photon number distribution at the output, and calculated the frequency shift and its variance, based on the definitions in Eq.~(\ref{eq20}) and (\ref{eq38}). However, there should be vacuum fluctuations of an electromagnetic field, which produce fluctuations in the photon number and consequently quantum noises, i.e. shot noise and radiation pressure noise. To clarify the usefulness of the WVA, we need to appropriately take into account quantum noises and evaluate SNR. These will be done in the next section.

%%%%%%%%%%%%%%%%%%%%%%%%%%%%%%%%%%%%%
\section{Quantum noise and standard quantum limit}
\label{sec4} 

In this section, we first introduce the quantum fluctuations of photon number and compute shot noise and radiation pressure noise in a unified framework in the Heisenberg picture. Then defining SNR, we derive the SQL and the sensitivity limit to measure position of a mirror.

From Eqs.~(\ref{eq13}) and (\ref{eq23}), the fluctuating part of the photon number is
\begin{align}
\Delta n (\omega) &\equiv n(\omega)-\bar{n}(\omega) \nonumber \\
&= ( \alpha^{*} d + \alpha d^{\dag}) \sin^2 \omega \xi \nonumber \\
&+\left\{ 2 |\alpha|^2  \omega \xi_{\rm{r}} + i ( \alpha a^{\dag} - \alpha^{*} a ) \right\} \sin \omega \xi \cos \omega \xi \;. \label{eq34}
\end{align}
Then an additional contribution to the frequency shift due to a deviation from the average photon number is given by 
\begin{equation}
\Delta \tilde{\Omega} \approx \omega_0 \int_{-\infty}^{\infty} \frac{d\tilde{\Omega}}{2\pi} \, \tilde{\Omega}\,\frac{\Delta n(\tilde{\Omega})}{N} \nonumber \\
\end{equation}
The expectation value of $\Delta \tilde{\Omega}$ is zero by definition. The quantum noise arises from the variance of $\Delta \tilde{\Omega}$, that is, $\langle (\Delta \tilde{\Omega})^2 \rangle$.

Since in our case each frequency mode of a laser pulse is independent when evaluated in a vacuum state $| 0 \rangle \equiv |0\rangle_d |0\rangle_a$, we have    
\begin{align}
\left\langle \{ \alpha d^{\dag} + \alpha^{*} d\}\{\alpha^{\prime} ( d^{\dag})^{\prime} + (\alpha^{*})^{\prime} d^{\prime}\} \right\rangle &= 2\pi \bar{n}_0(\omega) \delta (\omega-\omega^{\prime}) \;, \label{eq27} \\
\left\langle (\alpha a^{\dag} - \alpha^{*} a) (\alpha^{\prime} (a^{\dag})^{\prime} - (\alpha^{*})^{\prime} a^{\prime}) \right\rangle &= - 2\pi \bar{n}_0(\omega) \delta (\omega-\omega^{\prime})  \;.\label{eq28} 
\end{align}
One needs to be careful about that the vacuum field $a$ and that in $\xi_{\rm{r}}$ are defined at different times, because $\xi_{\rm{r}}$ is induced by a laser pulse at the past time. To distinguish them, we fix the subscript "1" for the vacuum field at the time of the previous laser pulse. Then using Eq.~(\ref{eq46}), $\xi_{\rm{r}}^2$ is given by
\begin{align}
\langle \xi_{\rm{r}}^2 \rangle &= \left(\frac{2T}{m}\right)^2 \left\langle \int_0^{\infty} \frac{d\omega}{2\pi} \, \hbar \omega (\alpha a_1^{\dag} + \alpha^{*} a_1) \right. \nonumber \\
& \left. \quad \quad \times \int_0^{\infty} \frac{d\omega^{\prime}}{2\pi} \, \hbar \omega^{\prime} (\alpha^{\prime} (a_1^{\prime})^{\dag} + (\alpha^{\prime})^{*} a_1^{\prime}) \right\rangle  \nonumber \\
&= \left(\frac{2T}{m}\right)^2 \int_0^{\infty} \frac{d\omega}{2\pi} \, (\hbar \omega)^2 \bar{n}_0(\omega) \nonumber \\
&\approx \left(\frac{2 \hbar \omega_0 T}{m}\right)^2 (1+\tilde{\sigma}_{\omega}^2) N_0 \;. \label{eq16} 
\end{align}
Here we extended the integral range from $-\omega_0$ to $-\infty$ owing to the exponential suppression and used Eq.~(\ref{eq26}). As we prove below, it is shown that there is no correlation between vacuum fields at different times. The correlation term between the vacuum fields in the same frequency mode but at different times by $T$ is
\begin{align}
&\left\langle \left\{ \alpha a_1^{\dag} + \alpha^{*} a_1 \right\} \left\{ \alpha a^{\dag} - \alpha^{*} a \right\} \right. \nonumber \\
& \left. +\left\{ \alpha a^{\dag} - \alpha^{*} a \right\} \left\{ \alpha a_1^{\dag} + \alpha^{*} a_1 \right\} \right\rangle \nonumber \\
&=\left\langle \left\{ \alpha a^{\dag} e^{-i \omega T} + \alpha^{*} a\, e^{i \omega T}\right\} \left\{ \alpha a^{\dag} - \alpha^{*} a \right\} \right. \nonumber \\
& \left. + \left\{ \alpha a^{\dag} - \alpha^{*} a \right\} \left\{ \alpha a^{\dag} e^{-i \omega T} + \alpha^{*} a\, e^{i \omega T}\right\} \right\rangle \nonumber \\
&= 2i\, |\alpha|^2 \langle  a a^{\dag} \rangle \sin \omega T \;. \nonumber 
\end{align}
Actually there exists a nonzero correlation, which is proportional to $\sin \omega T$. However, in practice, the frequency resolution of a photodetector (multichannel CCD) is finite and a single frequency bin contains a number of frequency modes. So summing over the modes averages out the phase of the correlation term. This is a conspicuous contrast to a continuous measurement, in which the light spectrum is nearly monochromatic and the correlation term almost coherently oscillates, with the phase difference of the order of $\tilde{\Omega} T \ll 1$. Thus, we conclude that we can treat the fields $a$ and $a_1$ independently in the pulsed measurements. 

Therefore, from Eq.~(\ref{eq34}) and the above relations, we obtain
\begin{align}
\langle \Delta n(\Omega)\Delta n(\Omega^{\prime}) \rangle &= 2\pi \bar{n}_0(\omega) \sin^2 \omega \xi \, \delta (\omega-\omega^{\prime}) \nonumber \\
&+ \langle \xi_r^2 \rangle \bar{n}_0(\omega) \bar{n}_0(\omega^{\prime}) \omega \omega^{\prime} \sin 2 \omega \xi \sin 2 \omega^{\prime} \xi \nonumber \\
&= 2\pi \frac{\bar{n}(\tilde{\Omega})}{\omega_0} \delta (\tilde{\Omega}-\tilde{\Omega}^{\prime}) + \omega_0^2 \langle \xi_r^2 \rangle \beta(\tilde{\Omega})\beta(\tilde{\Omega}^{\prime}) \;,
\label{eq29}
\end{align}
with
\begin{equation}
\beta(\tilde{\Omega}) \equiv \bar{n}_0(\tilde{\Omega}) (1+\tilde{\Omega})  \sin [(1+\tilde{\Omega})\phi + \theta] \;.
\end{equation}
The first term is shot noise, which results from the Poissonian statistics of photons and recovers Eq.~(7) in \cite{Nishizawa:2012weak}. The second term is radiation pressure noise that we originally derived here in the context of the weak value amplification in the Heisenberg picture. Then the variance of the frequency shift is
\begin{align}
\langle (\Delta \tilde{\Omega})^2 \rangle &= \frac{\omega_0^2}{N^2} \int_{-\infty}^{\infty} \frac{d\tilde{\Omega}}{2\pi} \int_{-\infty}^{\infty} \frac{d\tilde{\Omega}^{\prime}}{2\pi} \,\tilde{\Omega} \tilde{\Omega}^{\prime} \langle \Delta n(\tilde{\Omega}) \Delta n(\tilde{\Omega}^{\prime}) \rangle \nonumber \\
&\approx \frac{\langle \tilde{\Omega}^2 \rangle}{N} + \frac{P_0^2\tilde{\sigma}_{\omega}^4}{N^2\hbar^2} \langle \xi_{\rm{r}}^2 \rangle \sin^2 \theta \nonumber \\
&\approx \frac{\langle (\Delta \tilde{\Omega})^2 \rangle_{\rm{SQL}}}{2} \left[ \frac{1}{I} + I \right] \;. 
\label{eq51}
\end{align}  
At the second line, we assumed a weak measurement ($|\phi| \ll 1$) and kept the leading order term in powers of small $\phi$. At the third line, we used Eq.~(\ref{eq16}) and the approximated photon number from Eq.~(\ref{eq15})
\begin{equation}
N \approx N_0 \sin^2 \frac{\theta}{2}  \label{eq44} \;,
\end{equation}
and defined
\begin{align}
I &\equiv N_0 \eta \label{eq47} \;, \\
\eta &\equiv \frac{4\hbar \omega_0 \sigma_{\omega} T}{m} \left| \cos \frac{\theta}{2} \right| \sqrt{1+ \tilde{\sigma}_{\omega}^2} \;, \\
\langle (\Delta \tilde{\Omega})^2 \rangle_{\rm{SQL}} 
&\equiv 2 \tilde{\sigma}_{\omega}^2 \eta \left| \sin \frac{\theta}{2} \right|^{-2} \nonumber \\
&= \frac{2 \tilde{\sigma}_{\omega}^2 \eta}{|\cos (\theta/2)|^2} \left| A_w \right|^2 \;. \label{eq35}
\end{align}
These parameters, $I$ and $\eta$, are interpreted as the intensity of a measurement and the susceptibility of a mirror to measurement disturbance, respectively. We see in the variance that the first term is shot noise and the second term is radiation pressure noise and that there is a tradeoff between these two terms, which consequently gives the lower bound of total noise that one can reach. The minimum total noise is achievable when $I=1$, for which the noise is $\langle (\Delta \tilde{\Omega})^2 \rangle_{\rm{SQL}}$. This sensitivity limit is the so-called SQL. 

To derive minimum mirror displacement that is detectable by pulsed measurements, we define the SNR for a pointer shift:
\begin{align}
{\rm{SNR}} \equiv \frac{|\langle \tilde{\Omega} \rangle|}{\sqrt{\langle \Delta \tilde{\Omega}^2 \rangle}} \;. \nonumber
\end{align}
Since the shift signal in Eq.~(\ref{eq20}) is expanded for small $\phi$ as 
\begin{equation}
\left| \langle \tilde{\Omega} \rangle \right| \approx \tilde{\sigma}_{\omega}^2 \left| \phi \cot \frac{\theta}{2} \right| = \tilde{\sigma}_{\omega}^2 | \phi A_w| \;, \label{eq49}
\end{equation}
the optimal SNR is
\begin{equation}
{\rm{SNR}}_{\rm{SQL}}^2 = \frac{m}{8T \hbar \omega_0^2} \left( \frac{\tilde{\sigma}_{\omega}^2}{1+ \tilde{\sigma}_{\omega}^2} \right)^{1/2} \left| \cos \frac{\theta}{2} \right| |\phi|^2 \nonumber \\ 
\end{equation}
Using $|\phi|=4\omega_0 \ell$ and setting ${\rm{SNR}}=1$, we obtain the minimally detectable mirror displacement is 
\begin{align}
\ell_{\rm{SQL}} &= \left(1+ \frac{1}{\tilde{\sigma}_{\omega}^2}\right)^{1/4} \sqrt{\frac{ T \hbar}{2m}}
 \left| \cos \frac{\theta}{2} \right|^{-1/2} \nonumber \\
&\geq \sqrt{\frac{ T \hbar}{2m}} \;. \label{eq33}
\end{align}
This exactly coincides with the standard quantum limit derived from an elementary argument on the SQL based on the Heisenberg uncertainty relation in two-time position measurements \cite{Braginsky:book}. This sensitivity limit does not depend on the laser power of an experiment and gives the fundamental limit imposed by the Heisenberg inequality. To increase the sensitivity further, one can choose shorter time interval $T$ of a measurement. However, one needs to increase $N_0$ to compensate the decrease in $T$ and keep $I=1$ ($N_0 T={\rm{const.}}$). Namely, to improve the sensitivity to $\ell$ by 10 times, one needs $10^2$ times larger laser power. This is the standard scaling appearing in quantum metrology when photons in each frequency are not correlated \cite{Giovannetti:2004rev}.

The interesting result here is that there is no amplification of the SNR due to the WVA. Indeed, both the signal and the noise are amplified. However, in the SQL formula, both amplifications are canceled out and the weak value disappears. This can be understood intuitively from the lower plot in Fig.~\ref{fig1a}. As $\theta$ decreases (the weak value is more amplified), the distribution is suppressed and becomes less sharp. More strictly speaking, the shot noise is amplified due to the small statistics at the output and the radiation pressure noise is amplified by the same mechanism as the signal because these come from the displacements of a mirror. Therefore, not only signal but also quantum noise are amplified. If the distribution is normalized by $N$ as in Fig.~\ref{fig1b}, one sees only signal amplification, which often causes the confusion in that the WVA is useful for improving sensitivity. 

We comment on the experimental feasibility of the SQL in a pulsed measurement. The SQL is given for typical parameters by 
\begin{equation}
\ell_{\rm{SQL}} \geq 7.3\times 10^{-18} \left( \frac{T}{10^{-3}\,{\rm{s}}} \right)^{1/2} \left( \frac{1.0\,{\rm{g}}}{m} \right)^{1/2} \, {\rm{m}} \;. 
\end{equation}
To reach the SQL, we need $I=1$. For typical parameters, $I$ is written as 
\begin{align}
I &\approx 1.1 \times \left( \frac{1.0\,{\rm{g}}}{m} \right) \left( \frac{\omega_0}{1.8 \times 10^{15}\,{\rm{s}}^{-1}} \right) \left( \frac{T}{10^{-3}\,{\rm{s}}} \right) \left( \frac{P_0}{10\,{\rm{J}}} \right) \nonumber \\
&\times \left( \frac{1+\tilde{\sigma}_{\omega}^2}{2} \right)^{1/2} \tilde{\sigma}_{\omega} \cos \frac{\theta}{2} \;.
\end{align}
Since the requirement for these parameters is not so severe, the order of unity for $I$ and the SQL are achievable with state-of-art technologies.

%%%%%%%%%%%%%%%%%%%%%%%%%%%%%%%%%%%%%%%%%
\section{Quantum technique to overcome the SQL}
\label{sec5} 

There are many methods to overcome the SQL in a standard quantum metrology, such as using quantum entanglement and a squeezed state. These quantum correlations are also applicable to weak measurements with weak value amplification and enhance the sensitivity of parameter estimation \cite{Pang2014a,Pang2014b}. However, what the authors considered was not an optical interferometry with a bunch of photons but just multi-time measurements and is not relevant to our case considered in this paper. On the other hand, in some cases, say, in the measurement of a coupling parameter between a spin-1/2 particle and coherent light, it is shown that the Heisenberg scaling can be achieved with classical coherent light \cite{Zhang2013arXiv,Jordan2014QSMF}, though this does not necessarily mean that a weak measurement with a post-selection outperforms standard interferometry.

In this section, we consider one of examples that circumvent the SQL in optical pulsed measurements, namely, the squeezed vacuum input. This is implemented by replacing a coherent vacuum injected from the output port with a squeezed vacuum. That is, the output photon distribution is evaluated in the state 
\begin{equation}
|0_s \rangle = |0\rangle_d \otimes |0_s\rangle_a  \;,
\end{equation}
where $|0_s\rangle_a$ is the squeezed vacuum for the field $a$ at each frequency and is defined by
\begin{equation}
|0_s\rangle_a \equiv S(r_s,\phi_s) | 0 \rangle_a \;, 
\end{equation}
with the squeezing operator
\begin{align}
S(r_s,\phi_s) &\equiv \exp \left[ \int \frac{d\omega}{2\pi} \frac{r_s(\omega)}{2} \right. \nonumber \\
& \left.  \times \left\{ a^2(\omega) e^{-i\phi_s(\omega)} - \left[a^{\dag}(\omega) \right]^2 
e^{i\phi_s(\omega)} \right\} \right] \;. 
\end{align}
Note that the squeezing factor $r_s$ and angle $\phi_s$ depend on frequency. However, for simplicity of notation, we omit the arguments of $r_s$ and $\phi_s$ in what follows. The annihilation and creation operators are converted by the squeezing operator as
\begin{align}
S^{\dag}(r_s,\phi_s) a S(r_s,\phi_s) &= a \cosh r_s - a^{\dag} e^{i\phi_s} \sinh r_s \;, \\
S^{\dag}(r_s,\phi_s) a^{\dag} S(r_s,\phi_s) &= a^{\dag} \cosh r_s - a\, e^{-i\phi_s} \sinh r_s \;.
\end{align}
Since the squeezing operator does not change the linearity of $a$ and $a^{\dag}$, we have
\begin{equation}
_a\langle 0_s | a | 0_s \rangle_a =0 \quad \quad _a\langle 0_s | a^{\dag} | 0_s \rangle_a =0 \;.  
\end{equation}
Then the quantities that is linear in $a$ and $a^{\dag}$, e.g. $\bar{n}$, $N$, $\Delta n (\omega)$, $\langle \tilde{\Omega} \rangle$, and $\langle \tilde{\Omega}^2 \rangle$, are the same as those in the coherent vacuum case. However, the nonlinear terms in $a$ and $a^{\dag}$, that is, shot noise and radiation pressure noise, are modified. 

The derivation of quantum noise is provided in Appendix \ref{app:noise-in-squeezing} and the final expression when $|\phi| \ll 1$ is given by
\begin{align}
\langle (\Delta \tilde{\Omega})^2 \rangle_s &= \frac{\langle (\Delta \tilde{\Omega})^2 \rangle_{\rm{SQL}}}{2} \nonumber \\
& \times \left[ \frac{1+F_{+}(r_{s2},\phi_{s2})}{I} + I \left\{1+f_- (r_{s1},\phi_{s1}) \right\} \right] \;,
\label{eq48}
\end{align}
with
\begin{align}
f_{\pm} (r_s,\phi_s) &= 2 \sinh r_s (\sinh r_s \pm \cos \phi_s \cosh r_s)\;, \nonumber \\
F_{+} (r_{s}, \phi_{s}) &= f_{+}(r_{s},\phi_{s}) \cos^2 \frac{\theta}{2} \;. \nonumber  
\end{align}
In the above equations, $I$ is the same one as in Eq.~(\ref{eq47}). We fixed the subscripts "1" and "2" to denote the first and the second laser pulses because $r_s$ and $\phi_s$ can take different values at different times. However, to obtain the analytic expression of shot noise and radiation pressure noise, we assumed that the squeezing factor and squeezing angle are constant and independent of frequency. Of course, in practice, the squeezing factor and angle depend on frequency and are not reduced to the simple expression in Eq.~(\ref{eq48}). In that case, the quantum noise spectrum is somewhat degraded and becomes closer to the conventional spectrum without the squeezing. Since the vacuum fields at different times are not correlated due to a broadband spectrum as in the case of a coherent vacuum, there is again no correlation term between shot noise and radiation pressure noise. 

\begin{figure}[h]
\begin{center}
\includegraphics[width=8.5cm]{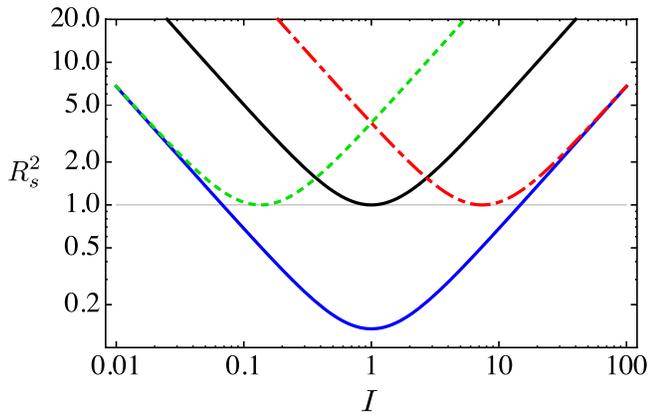}
\caption{Ratio of quantum noise $R_s^2$ with/without the squeezed vacuum input. The black, solid curve is conventional case ($r_s=0$). The other curves are when the squeezing factor is set to $r_s=1$ and the squeezing angles are $(\phi_{s1},\phi_{s2})=(0,0)$ (green, dotted), $(\pi,\pi)$ (red, dotted-dashed), and $(0,\pi)$ (blue, solid).}
\label{fig3}
\end{center}
\end{figure}

Since the frequency shift in Eq.~(\ref{eq49}) is the same as in the coherent vacuum case and does not depend on the squeezing parameters, it is convenient to define the ratio of noise in the squeezed vacuum case to that in the coherent vacuum case,
\begin{equation}
R_{\rm{s}}^2 \equiv \frac{\langle (\Delta \tilde{\Omega})^2 \rangle_s}{\langle (\Delta \tilde{\Omega})^2 \rangle_{\rm{SQL}}} \;.
\end{equation}
For specific choices of the squeezing angles,
\begin{align}
f_{\pm} (r_s,0) &= \pm 2 \sinh r_s e^{\pm r_s} \quad \longrightarrow \quad 1+f_{\pm} = e^{\pm 2r_s} \;, \\
f_{\pm} (r_s,\pi) &= \mp 2 \sinh r_s e^{\mp r_s} \quad \longrightarrow \quad  1+f_{\pm} = e^{\mp 2r_s} \;.
\end{align}
If $\theta$ is small, $F_+ \approx f_+$. Then for a set of squeezing parameters, $r_s= r_{s1}=r_{s2}$ and $\phi_{s} = \phi_{s1}=\phi_{s2}=0$, 
\begin{equation}
R_{\rm{s}}^2 = \frac{1}{2} \left[ \frac{e^{+2r_s}}{I} + I \, e^{-2r_s} \right] \;,
\end{equation}
and
for $r_s = r_{s1}=r_{s2}$ and $\phi_{s} = \phi_{s1}=\phi_{s2}=\pi$,
\begin{equation}
R_{\rm{s}}^2 = \frac{1}{2} \left[ \frac{e^{-2r_s}}{I} + I \, e^{+2r_s} \right] \;.
\end{equation}
In these cases, one of the shot noise or radiation pressure noise are reduced and the other is enhanced. This means that the balance between the shot noise or radiation pressure noise is change and then the optimal laser power is also changed. However, the quantum noise never circumvents the SQL. If one choose the squeezing angle optimally as $\phi_{s1}=0$ and $\phi_{s2}=\pi$, the noise is minimized: 
\begin{equation}
R_{\rm{s}}^2 = \frac{e^{-2r_s}}{2} \left[ \frac{1}{I} + I \right] \;,
\end{equation}
Therefore, we can overcome the SQL by a factor of $e^{-r_s}$ in SNR. In Fig.~\ref{fig3}, we show quantum noises as a function of $I$ for different choices of the squeezing angle.

%%%%%%%%%%%%%%%%%%%%%%%%%%%%%%%%%%%%%
\section{Conclusions}
\label{sec6}

We have studied the WVA in position measurements in an optical interferometer by appropriately taking account quantum shot noise and radiation pressure noise in the Heisenberg picture. The Heisenberg formulation enables us to intuitively understand what happens when the measurement outcome is post-selected and the weak value is amplified. As discussed in Sec.~\ref{sec3}, the photon number distribution as a function of frequency is shifted by weak value amplification but is significantly suppressed due to the post-selection. However, once this distribution is normalized by the total number of photons, the suppression of the statistics disappears. Therefore, it looks like that a signal is amplified without any cost and often leads to a wrong conclusion. However, working in the Heisenberg picture makes this type of confusion clear.  

Then we have took into account vacuum fluctuations of an electromagnetic field and derived quantum noises, i.e. shot noise and radiation pressure noise, in Eq.~(\ref{eq51}). By defining the SNR, we have shown that the sensitivity of a mirror position measurement is limited by the SQL in Eq.~(\ref{eq33}), which is the same as that in a standard interferometry. Interestingly, the signal in Eq.~(\ref{eq49}) is amplified by the WVA but the quantum noise Eq.~(\ref{eq35}) is also amplified. As a result, there is no amplification of SNR because in the SNR formula both amplifications are canceled out and the weak value disappears. We have also shown that the SQL can be overcome by implementing the squeezed vacuum input. 

Thus, we conclude that the WVA has no advantage to improve the sensitivity of position measurements when quantum noise dominates and that other quantum techniques must be combined with the WVA to circumvent the SQL.

%%%%%%%%%%%%%%%%%%%%%%%%%%%%%%%%%%%%%
\begin{acknowledgments}
A. N. is supported by JSPS Postdoctoral Fellowships for Research Abroad. A. N. thanks N.-K. Fujimoto and K. Nakamura for fruitful discussions at the early stage of this work, and also thank Yanbei Chen and Yiqiu Ma for carefully reading the manuscript and giving comments. 
\end{acknowledgments}

%%%%%%%%%%%%%%%%%%%%%%%%%%%%%%%%%%%%%%
\appendix

%%%%%%%%%%%%%%%%%%%%%%%%%%%%%%%%%%%%%%%%%%%%
\section{Input - output relation}
\label{app:io-relation}

\begin{figure}[h]
\begin{center}
\includegraphics[width=8cm]{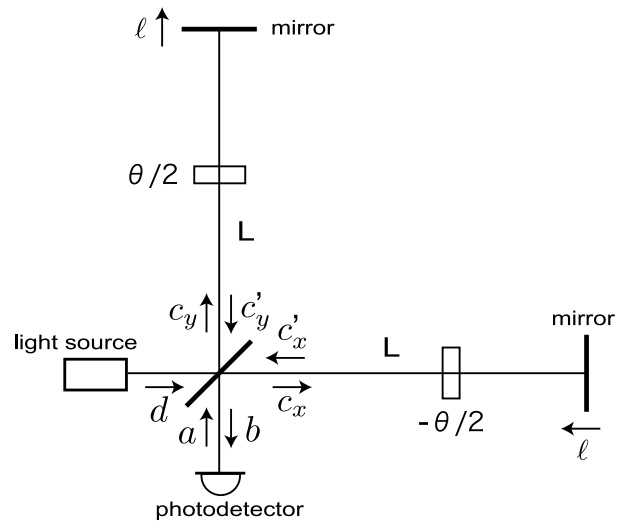}
\caption{Michelson interferometer and electric fields. Here the initial phase offset $\theta$ is symmetrized merely for simplicity of calculation.}
\label{fig4a}
\end{center}
\end{figure}

We derive an optical input-output relation in a Michelson interferometer shown in Fig.~\ref{fig4a}. 

\subsection{Field relation at the beam splitter}
At the beam splitter, the output field $b$ is related with the other fields $c_x^{'}$ and $c_y^{'}$ by
\begin{equation}
E_b (t) = \frac{1}{\sqrt{2}} \left[ E_{c_y^{'}}(t) - E_{c_x^{'}} (t) \right] \;. \nonumber
\end{equation}
Then using Eqs.~(\ref{eq22}) - (\ref{eq25}), we have the relation  
\begin{equation}
b= \frac{1}{\sqrt{2}} \left( c_y^{'} - c_x^{'} \right) \;.
\label{eq16a}
\end{equation}
On the other hand, the fields $c_x$, $c_y$ are expressed in terms of the input fields $D$ and $a$ as
\begin{equation}
c_y=\frac{1}{\sqrt{2}} \left( D + a \right) \;, \quad c_x=\frac{1}{\sqrt{2}} \left( D - a \right) \;. \label{eq17a}
\end{equation}

\subsection{Field relation in the arms}

When light goes around an interferometer arm whose length is $L$, the light takes time of a round trip in each arm, $2\tau \equiv 2L/c$, the short time delay due to small displacement of a mirror, $2(\ell+\delta \ell)/c$, and a phase shifter, $\Delta t_{\theta}$ for a round trip. The mirror displacement $\ell$ is defined as differential component of mirror displacement between both arms because a common mode can always be included in the definition of $L$. The difference of the definitions between $\ell$ and $\delta \ell$ is that the former is the mirror displacement without radiation pressure and the latter is due to radiation pressure noise coming from vacuum field fluctuations. Here $\Delta t_{\theta}$ is defined by the phase factor $\theta/2 \equiv \omega \, \Delta t_{\theta}(\omega)$ and the initial phase offset $\theta$ between two arms is symmetrized merely for simplicity of calculation. For the other arm, the signs of $\ell$, $\delta \ell$, and $\Delta t_{\theta}$ are reversed.

Denoting the time delays as 
\begin{equation}
\xi \equiv 2\ell + \Delta t_{\theta} \;, \quad \xi_{\rm{r}} \equiv 2 \delta \ell\;,  \nonumber
\end{equation}  
the electromagnetic fields departing from the beam splitter and returning there after the reflections at the mirrors are connected by the relations  
\begin{align}
E_{c^{'}_y} (t) &= E_{c_y} [ t-2\tau-(\xi+\xi_{\rm{r}})] \;, \nonumber \\ 
E_{c^{'}_x} (t) &= E_{c_x} [ t-2\tau+(\xi+\xi_{\rm{r}}) ] \;. \nonumber 
\end{align}
Note that $\xi_{\rm{r}}$ is an Hermitian operator. Then using Eqs.~(\ref{eq22}) - (\ref{eq25}), we hav the relation
\begin{equation}
c^{'}=c\,e^{i \omega (2\tau \pm \xi \pm \xi_{\rm{r}})} \;,
\label{eq36}
\end{equation}
where the upper and lower signs correspond to $y$ and $x$ arms, respectively.

\subsection{Radiation pressure noise}

A pulsed light in Fig.~\ref{fig4a} is reflected at the mirror. The number distributions of photons impinging on the mirror in each arm of an interferometer are given by
\begin{align}
n_y &\equiv c_y^{\dag} c_y = \frac{1}{2} (D^{\dag}D + a^{\dag}D + D^{\dag}a) \;, \nonumber \\
n_x &\equiv c_x^{\dag} c_x = \frac{1}{2} (D^{\dag}D - a^{\dag}D - D^{\dag}a) \;. \nonumber  
\end{align}
The number of photons included in a pulsed light is always fluctuating. In the following, we ignore the higher order terms in vacuum fluctuations and keep up to the first-order terms in the amplitude of vacuum fluctuations. Defining $D=\alpha + d$ by separating a classical part $\alpha$ and a quantum part $d$, we have the average photon number $\bar{n}_x=\bar{n}_y=|\alpha|^2/2$ and their fluctuations
\begin{align}
\Delta n_y &\equiv n_y - \bar{n}_y \nonumber \\
&= \frac{1}{2} (\alpha d^{\dag} + \alpha^{*} d + \alpha a^{\dag} + \alpha^{*} a) \;, \nonumber \\
\Delta n_x &\equiv n_x - \bar{n}_x \nonumber \\
&= \frac{1}{2} (\alpha d^{\dag} + \alpha^{*} d - \alpha a^{\dag} - \alpha^{*} a) \;. \nonumber 
\end{align}
Since the classical part can be compensated exactly by feedback controls, we do not consider it hereafter. The momentum fluctuation exerted on the mirror per a pulsed light (twice due to reflection) is 
\begin{equation}
\Delta P_i = 2\int_{0}^{\infty} \frac{d\omega}{2\pi} \hbar \omega \Delta n_i(\omega)\;, \quad i=x,y \;.\label{eq31}
\end{equation}

Let a pulse interval be $T$. Then at the next measurement time after the interval $T$, the mirror changes its position by
\begin{align}
\xi_{\rm{r}} &= \pm 2 \delta \ell \nonumber \\
&= \frac{2 T\Delta P_{y,x}}{m} \nonumber \\
&= \pm \frac{2 T}{m} \int_{0}^{\infty} \frac{d\omega}{2\pi} \hbar \omega (\alpha a^{\dag} + \alpha^{*} a) \;. \label{eq12} 
\end{align}
The upper sign is for the $y$ arm and the lower sign is for the $x$ arm. Also we neglect the common mode because it can always be included in the definition of $L$ without loss of generality.

\subsection{Electromagnetic field at the output}
\label{sec:output-EM-field}
Combining Eqs.~(\ref{eq16a}), (\ref{eq17a}), and (\ref{eq36}), the output field is
\begin{equation}
b = e^{2i \omega \tau} \left\{ i D \sin \omega (\xi_r + \xi ) + a \cos \omega (\xi_r + \xi ) \right\}\;. \nonumber 
\end{equation} 
In the following, we assume $|\omega \xi_r| \ll 1$. We also assumed that $a$ is a vacuum field
so that a higher order term $a \times \xi_r$ can be neglected. Thus, plugging in $D=\alpha + d$ and neglecting the higher order terms, we obtain 
\begin{equation}
b \approx i e^{2i \omega \tau} \left\{ \alpha \sin \omega \xi + \alpha \omega \xi_r \cos \omega \xi  + d\sin \omega \xi -i a \cos \omega \xi \right\} \;. 
\end{equation}

%%%%%%%%%%%%%%%%%%%%%%%%%%%%%%%%%%%%%%%%%%%%%
\section{Integral formulas}
\label{secC}

\begin{align}
&\int_{-\infty}^{\infty} e^{-ax^2} \cos (2bx+c) dx =\sqrt{\frac{\pi}{a}} e^{-b^2/a} \cos c \;, 
\label{eqc1} \\
&\int_{-\infty}^{\infty} e^{-ax^2} \sin (2bx+c) dx =\sqrt{\frac{\pi}{a}} e^{-b^2/a} \sin c 
\label{eqc2}\;, 
\end{align}

%%%%%%%%%%%%%%%%%%%%%%%%%%%%%%%%%%%%%%%%%%%%%
\section{Derivation of quantum noise in the case of the squeezed vacuum input}
\label{app:noise-in-squeezing}

Since the squeezing operator does not change the linearity of $a$ and $a^{\dag}$, we have
\begin{equation}
_a\langle 0_s | a | 0_s \rangle_a =0 \quad \quad _a\langle 0_s | a^{\dag} | 0_s \rangle_a =0 \;.  
\end{equation}
Then the quantities that is linear in $a$ and $a^{\dag}$, e.g. $\bar{n}$, $N$, $\Delta n (\omega)$, $\langle \tilde{\Omega} \rangle$, and $\langle \tilde{\Omega}^2 \rangle$, are the same as those in the coherent vacuum case. However, the nonlinear terms in $a$ and $a^{\dag}$ are modified:
\begin{align}
_a\langle 0_s | a^{\dag} (a^{\prime})^{\dag} | 0_s \rangle_a &= - 2\pi \delta (\omega-\omega^{\prime} ) e^{-i \phi_s} \sinh r_s \cosh r_s \;, \nonumber \\
_a\langle 0_s | a a^{\prime} | 0_s \rangle_a &= - 2\pi \delta (\omega-\omega^{\prime} ) e^{i \phi_s} \sinh r_s \cosh r_s \;, \nonumber \\
_a\langle 0_s | a (a^{\prime})^{\dag} | 0_s \rangle_a &= 2\pi \delta (\omega-\omega^{\prime} ) \cosh^2 r_s \;, \nonumber \\
_a\langle 0_s | a^{\dag} a^{\prime} | 0_s \rangle_a &= 2\pi \delta (\omega-\omega^{\prime} ) \sinh^2 r_s \;. \nonumber 
\end{align}
The terms giving shot noise and radiation pressure noise in Eq.~(\ref{eq34}) are
\begin{align}
& _a\langle 0_s | \{ \alpha a^{\dag} \pm \alpha^{\ast} a \} \{ \alpha^{\prime} (a^{\prime})^{\dag} \pm (\alpha^{\prime})^{\ast} a^{\prime} \} | 0_s \rangle_a \nonumber \\
&= - 2\pi \delta (\omega-\omega^{\prime} ) \left\{ \mp \bar{n}_0(\omega) \cosh 2 r_s + {\rm{Re}} [ \alpha^2 e^{-i \phi_s} ] \sinh 2r_s \right\} \;, \nonumber 
\end{align}
and
\begin{align}
\langle \xi_r^2 \rangle_s &\equiv \, _{a_1}\langle 0_s | \xi_r^2 | 0_s \rangle_{a_1} \nonumber \\
&= \left(\frac{2T}{m}\right)^2 \int_0^{\infty} \frac{d\omega}{2\pi} \int_0^{\infty} \frac{d\omega^{\prime}}{2\pi} \, \hbar \omega \hbar \omega^{\prime}\, \nonumber \\
&\times _{a_1} \langle 0_s | \{ \alpha a_1^{\dag} + \alpha^{*} a_1\} \{ \alpha^{\prime} (a_1^{\prime})^{\dag} + (\alpha^{\prime})^{*} a_1^{\prime} \} | 0_s \rangle_{a_1}  \nonumber \\
&= \left(\frac{2T}{m}\right)^2 \int_0^{\infty} \frac{d\omega}{2\pi} (\hbar \omega)^2 \, \nonumber \\
& \times \left\{ |\alpha|^2 \cosh 2r_{s1} - {\rm{Re}}[ \alpha^2 e^{-i \phi_{s1}} ] \sinh 2r_{s1} \right\} \;. 
\end{align}
As in the case of a coherent vacuum, there is no correlation between vacuum fields at different time because of a broadband spectrum.

In the above, $\alpha^2$ always appear in the product $\alpha^2 e^{-i \phi_s}$. Without loss of generality, we can assume that $\alpha$ is real by absorbing the phase into $\phi_s$. So we write
\begin{equation}
\cos \phi_s = \frac{{\rm{Re}}[ \alpha^2 e^{-i \phi_s} ]}{|\alpha|^2} \;, \quad \quad \sin \phi_s = - \frac{{\rm{Im}}[ \alpha^2 e^{-i \phi_s} ]}{|\alpha|^2} \;. \nonumber
\end{equation}
In addition, we distinguish $r_s$ and $\phi_s$ at the times by fixing the subscripts "1" and "2" for the first and the second laser pulses. To obtain the analytic expression of shot noise and radiation pressure noise, we assume that the squeezing factor and squeezing angle are constant and independent of frequency. Combining all these and using Eq.~(\ref{eq34}), we obtain
\begin{widetext}
\begin{align}
\langle \Delta n(\omega)\Delta n(\omega^{\prime}) \rangle_s &= 2\pi \bar{n}_0(\omega) \sin^2 \omega \xi \, \delta (\omega-\omega^{\prime}) \left\{ 1+ f_{+}(r_{s2},\phi_{s2}) \cos^2 \omega \xi \right\} + \langle \xi_r^2 \rangle_s \bar{n}_0(\omega) \bar{n}_0(\omega^{\prime}) \omega \omega^{\prime} \sin 2 \omega \xi \sin 2 \omega^{\prime} \xi \nonumber \\
&= 2\pi \frac{\bar{n}(\tilde{\Omega})}{\omega_0} \delta (\tilde{\Omega}-\tilde{\Omega}^{\prime}) \left\{ 1+ f_{+}(r_{s2},\phi_{s2}) \cos^2 \left[ (1+\tilde{\Omega})\frac{\phi}{2} + \frac{\theta}{2} \right] \right\} + \langle \xi_r^2 \rangle_s \omega_0^2 \beta(\tilde{\Omega})\beta(\tilde{\Omega}^{\prime})  \nonumber
\end{align}
with
\begin{align}
f_{\pm} (r_s,\phi_s) &\equiv 2 \sinh r_s (\sinh r_s \pm \cos \phi_s \cosh r_s) \nonumber \\
\langle \xi_r^2 \rangle_s &= \langle \xi_r^2 \rangle \left\{ 1+ f_{-}(r_{s1},\phi_{s1}) \right\} \;.
\end{align}

Then an additional term to the shot noise is
\begin{align}
&\frac{\omega_0^2}{N^2} \int_{-\infty}^{\infty} \frac{d\tilde{\Omega}}{2\pi} \int_{-\infty}^{\infty} \frac{d\tilde{\Omega}^{\prime}}{2\pi} \,\tilde{\Omega} \tilde{\Omega}^{\prime} 2\pi \frac{\bar{n}(\tilde{\Omega})}{\omega_0} \delta (\tilde{\Omega}-\tilde{\Omega}^{\prime}) f_{+}(r_{s2},\phi_{s2}) \cos^2 \left[ (1+\tilde{\Omega})\frac{\phi}{2} + \frac{\theta}{2} \right]  \nonumber \\
&=\frac{\sqrt{2\pi} N_0}{4\tilde{\sigma}_{\omega} N^2} f_{+}(r_{s2},\phi_{s2}) \int_{-\infty}^{\infty} \frac{d\tilde{\Omega}}{2\pi} \,\tilde{\Omega}^2  e^{-\tilde{\Omega}^2/2 \tilde{\sigma}_{\omega}^2} \sin^2 \left[ (1+\tilde{\Omega} ) \phi + \theta \right]   \nonumber \\
&= \frac{\langle \tilde{\Omega}^2 \rangle}{N} F_{+} (r_{s2}, \phi_{s2}) \;, \nonumber 
\end{align}
\end{widetext}
where
\begin{align}
F_{+} (r_{s2}, \phi_{s2}) &\equiv \frac{\tilde{\sigma}_{\omega}^2}{8N\langle \tilde{\Omega}^2 \rangle} N_0 f_{+}(r_{s2},\phi_{s2}) \nonumber \\
&\times \left\{ 1- e^{-2 \phi^2 \tilde{\sigma}_{\omega}^2} (1-4 \phi^2 \tilde{\sigma}_{\omega}^2) \cos [2 (\phi +\theta)] \right\} \nonumber \\
&\approx f_{+}(r_{s2},\phi_{s2}) \cos^2 \frac{\theta}{2} \;. \nonumber  
\end{align}
At the second line, we expanded in small $\phi$ and took the leading order term. The radiation pressure term is simply given by replacing $\langle \xi_r^2 \rangle$ in the calculation of a coherent vacuum case with
\begin{equation}
\langle \xi_r^2 \rangle_s = \left\{ 1+ f_{-}(r_{s1},\phi_{s1}) \right\} \langle \xi_r^2 \rangle \;.
\end{equation}
Thus,
\begin{align}
\langle (\Delta \tilde{\Omega})^2 \rangle_s &= \frac{\omega_0^2}{N^2} \int_{-\infty}^{\infty} \frac{d\tilde{\Omega}}{2\pi} \int_{-\infty}^{\infty} \frac{d\tilde{\Omega}^{\prime}}{2\pi} \,\tilde{\Omega} \tilde{\Omega}^{\prime} \langle \Delta n(\tilde{\Omega}) \Delta n(\tilde{\Omega}^{\prime}) \rangle \nonumber \\
&\approx \frac{\langle \tilde{\Omega}^2 \rangle}{N} \left\{ 1+F_{+}(r_{s2},\phi_{s2}) \right\} \nonumber \\
& + \frac{N_0^2 \omega_0^2 \tilde{\sigma}_{\omega}^4}{N^2} \langle \xi_{\rm{r}}^2 \rangle \left\{ 1+ f_{-}(r_{s1},\phi_{s1}) \right\} \sin^2 \theta \nonumber \\
&\approx \frac{\langle (\Delta \tilde{\Omega})^2 \rangle_{\rm{SQL}}}{2} \nonumber \\
& \times \left[ \frac{1+F_{+}(r_{s2},\phi_{s2})}{I} + I \left\{1+f_- (r_{s1},\phi_{s1}) \right\} \right] \;.
\end{align}
At the third line, we used Eqs.~(\ref{eq44}) and (\ref{eq16}), expanded in small $\phi$.

%%%%%%%%%%%%%%%%%%%%%%%%%%%%%%%%%%%%%%%%%%%%%

\bibliography{/Volumes/USB-MEMORY/my-research/bibliography}

\begin{thebibliography}{42}%
\makeatletter
\providecommand \@ifxundefined [1]{%
 \@ifx{#1\undefined}
}%
\providecommand \@ifnum [1]{%
 \ifnum #1\expandafter \@firstoftwo
 \else \expandafter \@secondoftwo
 \fi
}%
\providecommand \@ifx [1]{%
 \ifx #1\expandafter \@firstoftwo
 \else \expandafter \@secondoftwo
 \fi
}%
\providecommand \natexlab [1]{#1}%
\providecommand \enquote  [1]{``#1''}%
\providecommand \bibnamefont  [1]{#1}%
\providecommand \bibfnamefont [1]{#1}%
\providecommand \citenamefont [1]{#1}%
\providecommand \href@noop [0]{\@secondoftwo}%
\providecommand \href [0]{\begingroup \@sanitize@url \@href}%
\providecommand \@href[1]{\@@startlink{#1}\@@href}%
\providecommand \@@href[1]{\endgroup#1\@@endlink}%
\providecommand \@sanitize@url [0]{\catcode `\\12\catcode `\$12\catcode
  `\&12\catcode `\#12\catcode `\^12\catcode `\_12\catcode `\%12\relax}%
\providecommand \@@startlink[1]{}%
\providecommand \@@endlink[0]{}%
\providecommand \url  [0]{\begingroup\@sanitize@url \@url }%
\providecommand \@url [1]{\endgroup\@href {#1}{\urlprefix }}%
\providecommand \urlprefix  [0]{URL }%
\providecommand \Eprint [0]{\href }%
\providecommand \doibase [0]{http://dx.doi.org/}%
\providecommand \selectlanguage [0]{\@gobble}%
\providecommand \bibinfo  [0]{\@secondoftwo}%
\providecommand \bibfield  [0]{\@secondoftwo}%
\providecommand \translation [1]{[#1]}%
\providecommand \BibitemOpen [0]{}%
\providecommand \bibitemStop [0]{}%
\providecommand \bibitemNoStop [0]{.\EOS\space}%
\providecommand \EOS [0]{\spacefactor3000\relax}%
\providecommand \BibitemShut  [1]{\csname bibitem#1\endcsname}%
\let\auto@bib@innerbib\@empty
%</preamble>
\bibitem [{\citenamefont {Aharonov}\ \emph {et~al.}(1988)\citenamefont
  {Aharonov}, \citenamefont {Albert},\ and\ \citenamefont
  {Vaidman}}]{Aharonov:1988xu}%
  \BibitemOpen
  \bibfield  {author} {\bibinfo {author} {\bibfnamefont {Y.}~\bibnamefont
  {Aharonov}}, \bibinfo {author} {\bibfnamefont {D.~Z.}\ \bibnamefont
  {Albert}}, \ and\ \bibinfo {author} {\bibfnamefont {L.}~\bibnamefont
  {Vaidman}},\ }\href {\doibase 10.1103/PhysRevLett.60.1351} {\bibfield
  {journal} {\bibinfo  {journal} {Phys.Rev.Lett.}\ }\textbf {\bibinfo {volume}
  {60}},\ \bibinfo {pages} {1351} (\bibinfo {year} {1988})}\BibitemShut
  {NoStop}%
%%CITATION = PRLTA,60,1351;%%
\bibitem [{\citenamefont {{Aharonov}}\ and\ \citenamefont
  {{Vaidman}}(2007)}]{Aharonov:2007LNP}%
  \BibitemOpen
  \bibfield  {author} {\bibinfo {author} {\bibfnamefont {Y.}~\bibnamefont
  {{Aharonov}}}\ and\ \bibinfo {author} {\bibfnamefont {L.}~\bibnamefont
  {{Vaidman}}},\ }in\ \href {\doibase 10.1007/978-3-540-73473-4_13} {\emph
  {\bibinfo {booktitle} {Lecture Notes in Physics, Berlin Springer Verlag}}},\
  \bibinfo {series} {Lecture Notes in Physics, Berlin Springer Verlag}, Vol.\
  \bibinfo {volume} {734},\ \bibinfo {editor} {edited by\ \bibinfo {editor}
  {\bibfnamefont {J.~G.}\ \bibnamefont {{Muga}}}, \bibinfo {editor}
  {\bibfnamefont {R.~S.}\ \bibnamefont {{Mayato}}}, \ and\ \bibinfo {editor}
  {\bibfnamefont {l.~L.}\ \bibnamefont {{Egusquiza}}}}\ (\bibinfo {year}
  {2007})\ p.\ \bibinfo {pages} {399}\BibitemShut {NoStop}%
\bibitem [{\citenamefont {{Ritchie}}\ \emph {et~al.}(1991)\citenamefont
  {{Ritchie}}, \citenamefont {{Story}},\ and\ \citenamefont
  {{Hulet}}}]{Ritchie:1991}%
  \BibitemOpen
  \bibfield  {author} {\bibinfo {author} {\bibfnamefont {N.~W.~M.}\
  \bibnamefont {{Ritchie}}}, \bibinfo {author} {\bibfnamefont {J.~G.}\
  \bibnamefont {{Story}}}, \ and\ \bibinfo {author} {\bibfnamefont {R.~G.}\
  \bibnamefont {{Hulet}}},\ }\href {\doibase 10.1103/PhysRevLett.66.1107}
  {\bibfield  {journal} {\bibinfo  {journal} {Physical Review Letters}\
  }\textbf {\bibinfo {volume} {66}},\ \bibinfo {pages} {1107} (\bibinfo {year}
  {1991})}\BibitemShut {NoStop}%
\bibitem [{\citenamefont {Pryde}\ \emph {et~al.}(2005)\citenamefont {Pryde},
  \citenamefont {O'Brien}, \citenamefont {White}, \citenamefont {Ralph},\ and\
  \citenamefont {Wiseman}}]{Pryde:2004zw}%
  \BibitemOpen
  \bibfield  {author} {\bibinfo {author} {\bibfnamefont {G.}~\bibnamefont
  {Pryde}}, \bibinfo {author} {\bibfnamefont {J.}~\bibnamefont {O'Brien}},
  \bibinfo {author} {\bibfnamefont {A.}~\bibnamefont {White}}, \bibinfo
  {author} {\bibfnamefont {T.}~\bibnamefont {Ralph}}, \ and\ \bibinfo {author}
  {\bibfnamefont {H.}~\bibnamefont {Wiseman}},\ }\href {\doibase
  10.1103/PhysRevLett.94.220405} {\bibfield  {journal} {\bibinfo  {journal}
  {Phys.Rev.Lett.}\ }\textbf {\bibinfo {volume} {94}},\ \bibinfo {pages}
  {220405} (\bibinfo {year} {2005})},\ \Eprint
  {http://arxiv.org/abs/quant-ph/0412204} {arXiv:quant-ph/0412204 [quant-ph]}
  \BibitemShut {NoStop}%
%%CITATION = QUANT-PH/0412204;%%
\bibitem [{\citenamefont {{Resch}}\ \emph {et~al.}(2004)\citenamefont
  {{Resch}}, \citenamefont {{Lundeen}},\ and\ \citenamefont
  {{Steinberg}}}]{Resch:2004PhLA}%
  \BibitemOpen
  \bibfield  {author} {\bibinfo {author} {\bibfnamefont {K.~J.}\ \bibnamefont
  {{Resch}}}, \bibinfo {author} {\bibfnamefont {J.~S.}\ \bibnamefont
  {{Lundeen}}}, \ and\ \bibinfo {author} {\bibfnamefont {A.~M.}\ \bibnamefont
  {{Steinberg}}},\ }\href {\doibase 10.1016/j.physleta.2004.02.042} {\bibfield
  {journal} {\bibinfo  {journal} {Physics Letters A}\ }\textbf {\bibinfo
  {volume} {324}},\ \bibinfo {pages} {125} (\bibinfo {year} {2004})},\ \Eprint
  {http://arxiv.org/abs/quant-ph/0310091} {quant-ph/0310091} \BibitemShut
  {NoStop}%
\bibitem [{\citenamefont {{Wang}}\ \emph {et~al.}(2006)\citenamefont {{Wang}},
  \citenamefont {{Sun}}, \citenamefont {{Zhang}}, \citenamefont {{Jian-Li}},
  \citenamefont {{Huang}},\ and\ \citenamefont {{Guo}}}]{Wang:2006PRA}%
  \BibitemOpen
  \bibfield  {author} {\bibinfo {author} {\bibfnamefont {Q.}~\bibnamefont
  {{Wang}}}, \bibinfo {author} {\bibfnamefont {F.-W.}\ \bibnamefont {{Sun}}},
  \bibinfo {author} {\bibfnamefont {Y.-S.}\ \bibnamefont {{Zhang}}}, \bibinfo
  {author} {\bibnamefont {{Jian-Li}}}, \bibinfo {author} {\bibfnamefont
  {Y.-F.}\ \bibnamefont {{Huang}}}, \ and\ \bibinfo {author} {\bibfnamefont
  {G.-C.}\ \bibnamefont {{Guo}}},\ }\href {\doibase 10.1103/PhysRevA.73.023814}
  {\bibfield  {journal} {\bibinfo  {journal} {Phys. Rev. A}\ }\textbf {\bibinfo
  {volume} {73}},\ \bibinfo {eid} {023814} (\bibinfo {year}
  {2006})}\BibitemShut {NoStop}%
\bibitem [{\citenamefont {{Hosten}}\ and\ \citenamefont
  {{Kwiat}}(2008)}]{Hosten:2008}%
  \BibitemOpen
  \bibfield  {author} {\bibinfo {author} {\bibfnamefont {O.}~\bibnamefont
  {{Hosten}}}\ and\ \bibinfo {author} {\bibfnamefont {P.}~\bibnamefont
  {{Kwiat}}},\ }\href {\doibase 10.1126/science.1152697} {\bibfield  {journal}
  {\bibinfo  {journal} {Science}\ }\textbf {\bibinfo {volume} {319}},\ \bibinfo
  {pages} {787} (\bibinfo {year} {2008})}\BibitemShut {NoStop}%
\bibitem [{\citenamefont {{Dixon}}\ \emph {et~al.}(2009)\citenamefont
  {{Dixon}}, \citenamefont {{Starling}}, \citenamefont {{Jordan}},\ and\
  \citenamefont {{Howell}}}]{Dixon:2009PRL}%
  \BibitemOpen
  \bibfield  {author} {\bibinfo {author} {\bibfnamefont {P.~B.}\ \bibnamefont
  {{Dixon}}}, \bibinfo {author} {\bibfnamefont {D.~J.}\ \bibnamefont
  {{Starling}}}, \bibinfo {author} {\bibfnamefont {A.~N.}\ \bibnamefont
  {{Jordan}}}, \ and\ \bibinfo {author} {\bibfnamefont {J.~C.}\ \bibnamefont
  {{Howell}}},\ }\href {\doibase 10.1103/PhysRevLett.102.173601} {\bibfield
  {journal} {\bibinfo  {journal} {Physical Review Letters}\ }\textbf {\bibinfo
  {volume} {102}},\ \bibinfo {eid} {173601} (\bibinfo {year} {2009})},\ \Eprint
  {http://arxiv.org/abs/0906.4828} {arXiv:0906.4828 [quant-ph]} \BibitemShut
  {NoStop}%
\bibitem [{\citenamefont {{Starling}}\ \emph {et~al.}(2009)\citenamefont
  {{Starling}}, \citenamefont {{Dixon}}, \citenamefont {{Jordan}},\ and\
  \citenamefont {{Howell}}}]{Starling:2009PRA}%
  \BibitemOpen
  \bibfield  {author} {\bibinfo {author} {\bibfnamefont {D.~J.}\ \bibnamefont
  {{Starling}}}, \bibinfo {author} {\bibfnamefont {P.~B.}\ \bibnamefont
  {{Dixon}}}, \bibinfo {author} {\bibfnamefont {A.~N.}\ \bibnamefont
  {{Jordan}}}, \ and\ \bibinfo {author} {\bibfnamefont {J.~C.}\ \bibnamefont
  {{Howell}}},\ }\href {\doibase 10.1103/PhysRevA.80.041803} {\bibfield
  {journal} {\bibinfo  {journal} {Phys. Rev. A}\ }\textbf {\bibinfo {volume}
  {80}},\ \bibinfo {eid} {041803} (\bibinfo {year} {2009})},\ \Eprint
  {http://arxiv.org/abs/0910.2410} {arXiv:0910.2410 [quant-ph]} \BibitemShut
  {NoStop}%
\bibitem [{\citenamefont {{Starling}}\ \emph {et~al.}(2010)\citenamefont
  {{Starling}}, \citenamefont {{Dixon}}, \citenamefont {{Williams}},
  \citenamefont {{Jordan}},\ and\ \citenamefont
  {{Howell}}}]{Starling:2010aPRA}%
  \BibitemOpen
  \bibfield  {author} {\bibinfo {author} {\bibfnamefont {D.~J.}\ \bibnamefont
  {{Starling}}}, \bibinfo {author} {\bibfnamefont {P.~B.}\ \bibnamefont
  {{Dixon}}}, \bibinfo {author} {\bibfnamefont {N.~S.}\ \bibnamefont
  {{Williams}}}, \bibinfo {author} {\bibfnamefont {A.~N.}\ \bibnamefont
  {{Jordan}}}, \ and\ \bibinfo {author} {\bibfnamefont {J.~C.}\ \bibnamefont
  {{Howell}}},\ }\href {\doibase 10.1103/PhysRevA.82.011802} {\bibfield
  {journal} {\bibinfo  {journal} {Phys. Rev. A}\ }\textbf {\bibinfo {volume}
  {82}},\ \bibinfo {eid} {011802} (\bibinfo {year} {2010})},\ \Eprint
  {http://arxiv.org/abs/0912.2357} {arXiv:0912.2357 [quant-ph]} \BibitemShut
  {NoStop}%
\bibitem [{\citenamefont {{Kofman}}\ \emph {et~al.}(2012)\citenamefont
  {{Kofman}}, \citenamefont {{Ashhab}},\ and\ \citenamefont
  {{Nori}}}]{Kofman:2012rev}%
  \BibitemOpen
  \bibfield  {author} {\bibinfo {author} {\bibfnamefont {A.~G.}\ \bibnamefont
  {{Kofman}}}, \bibinfo {author} {\bibfnamefont {S.}~\bibnamefont {{Ashhab}}},
  \ and\ \bibinfo {author} {\bibfnamefont {F.}~\bibnamefont {{Nori}}},\ }\href
  {\doibase 10.1016/j.physrep.2012.07.001} {\bibfield  {journal} {\bibinfo
  {journal} {Physics Report}\ }\textbf {\bibinfo {volume} {520}},\ \bibinfo
  {pages} {43} (\bibinfo {year} {2012})},\ \Eprint
  {http://arxiv.org/abs/1109.6315} {arXiv:1109.6315 [quant-ph]} \BibitemShut
  {NoStop}%
\bibitem [{\citenamefont {{Dressel}}\ \emph {et~al.}(2014)\citenamefont
  {{Dressel}}, \citenamefont {{Malik}}, \citenamefont {{Miatto}}, \citenamefont
  {{Jordan}},\ and\ \citenamefont {{Boyd}}}]{Dressel:2014rev}%
  \BibitemOpen
  \bibfield  {author} {\bibinfo {author} {\bibfnamefont {J.}~\bibnamefont
  {{Dressel}}}, \bibinfo {author} {\bibfnamefont {M.}~\bibnamefont {{Malik}}},
  \bibinfo {author} {\bibfnamefont {F.~M.}\ \bibnamefont {{Miatto}}}, \bibinfo
  {author} {\bibfnamefont {A.~N.}\ \bibnamefont {{Jordan}}}, \ and\ \bibinfo
  {author} {\bibfnamefont {R.~W.}\ \bibnamefont {{Boyd}}},\ }\href {\doibase
  10.1103/RevModPhys.86.307} {\bibfield  {journal} {\bibinfo  {journal}
  {Reviews of Modern Physics}\ }\textbf {\bibinfo {volume} {86}},\ \bibinfo
  {pages} {307} (\bibinfo {year} {2014})},\ \Eprint
  {http://arxiv.org/abs/1305.7154} {arXiv:1305.7154 [quant-ph]} \BibitemShut
  {NoStop}%
\bibitem [{\citenamefont {{Shikano}}(2011)}]{Shikano:2011}%
  \BibitemOpen
  \bibfield  {author} {\bibinfo {author} {\bibfnamefont {Y.}~\bibnamefont
  {{Shikano}}},\ }\href@noop {} {\bibfield  {journal} {\bibinfo  {journal}
  {ArXiv e-prints}\ } (\bibinfo {year} {2011})},\ \Eprint
  {http://arxiv.org/abs/1110.5055} {arXiv:1110.5055 [quant-ph]} \BibitemShut
  {NoStop}%
\bibitem [{\citenamefont {{Brunner}}\ and\ \citenamefont
  {{Simon}}(2010)}]{Brunner:2010}%
  \BibitemOpen
  \bibfield  {author} {\bibinfo {author} {\bibfnamefont {N.}~\bibnamefont
  {{Brunner}}}\ and\ \bibinfo {author} {\bibfnamefont {C.}~\bibnamefont
  {{Simon}}},\ }\href {\doibase 10.1103/PhysRevLett.105.010405} {\bibfield
  {journal} {\bibinfo  {journal} {Physical Review Letters}\ }\textbf {\bibinfo
  {volume} {105}},\ \bibinfo {eid} {010405} (\bibinfo {year} {2010})},\ \Eprint
  {http://arxiv.org/abs/0911.5139} {arXiv:0911.5139 [quant-ph]} \BibitemShut
  {NoStop}%
\bibitem [{\citenamefont {{Jordan}}\ \emph
  {et~al.}(2014{\natexlab{a}})\citenamefont {{Jordan}}, \citenamefont
  {{Mart{\'{\i}}nez-Rinc{\'o}n}},\ and\ \citenamefont
  {{Howell}}}]{Jordan:2014PRX}%
  \BibitemOpen
  \bibfield  {author} {\bibinfo {author} {\bibfnamefont {A.~N.}\ \bibnamefont
  {{Jordan}}}, \bibinfo {author} {\bibfnamefont {J.}~\bibnamefont
  {{Mart{\'{\i}}nez-Rinc{\'o}n}}}, \ and\ \bibinfo {author} {\bibfnamefont
  {J.~C.}\ \bibnamefont {{Howell}}},\ }\href {\doibase
  10.1103/PhysRevX.4.011031} {\bibfield  {journal} {\bibinfo  {journal}
  {Physical Review X}\ }\textbf {\bibinfo {volume} {4}},\ \bibinfo {eid}
  {011031} (\bibinfo {year} {2014}{\natexlab{a}})},\ \Eprint
  {http://arxiv.org/abs/1309.5011} {arXiv:1309.5011 [physics.optics]}
  \BibitemShut {NoStop}%
\bibitem [{\citenamefont {{Knee}}\ and\ \citenamefont
  {{Gauger}}(2014)}]{Knee2014PRX}%
  \BibitemOpen
  \bibfield  {author} {\bibinfo {author} {\bibfnamefont {G.~C.}\ \bibnamefont
  {{Knee}}}\ and\ \bibinfo {author} {\bibfnamefont {E.~M.}\ \bibnamefont
  {{Gauger}}},\ }\href {\doibase 10.1103/PhysRevX.4.011032} {\bibfield
  {journal} {\bibinfo  {journal} {Physical Review X}\ }\textbf {\bibinfo
  {volume} {4}},\ \bibinfo {eid} {011032} (\bibinfo {year} {2014})},\ \Eprint
  {http://arxiv.org/abs/1306.6321} {arXiv:1306.6321 [quant-ph]} \BibitemShut
  {NoStop}%
\bibitem [{\citenamefont {{di Lorenzo}}\ and\ \citenamefont
  {{Egues}}(2008)}]{Lorenzo:2008}%
  \BibitemOpen
  \bibfield  {author} {\bibinfo {author} {\bibfnamefont {A.}~\bibnamefont {{di
  Lorenzo}}}\ and\ \bibinfo {author} {\bibfnamefont {J.~C.}\ \bibnamefont
  {{Egues}}},\ }\href {\doibase 10.1103/PhysRevA.77.042108} {\bibfield
  {journal} {\bibinfo  {journal} {Phys. Rev. A}\ }\textbf {\bibinfo {volume}
  {77}},\ \bibinfo {eid} {042108} (\bibinfo {year} {2008})},\ \Eprint
  {http://arxiv.org/abs/0801.1814} {arXiv:0801.1814 [quant-ph]} \BibitemShut
  {NoStop}%
\bibitem [{\citenamefont {{Zhu}}\ \emph {et~al.}(2011)\citenamefont {{Zhu}},
  \citenamefont {{Zhang}}, \citenamefont {{Pang}}, \citenamefont {{Qiao}},
  \citenamefont {{Liu}},\ and\ \citenamefont {{Wu}}}]{Zhu:2011PRA}%
  \BibitemOpen
  \bibfield  {author} {\bibinfo {author} {\bibfnamefont {X.}~\bibnamefont
  {{Zhu}}}, \bibinfo {author} {\bibfnamefont {Y.}~\bibnamefont {{Zhang}}},
  \bibinfo {author} {\bibfnamefont {S.}~\bibnamefont {{Pang}}}, \bibinfo
  {author} {\bibfnamefont {C.}~\bibnamefont {{Qiao}}}, \bibinfo {author}
  {\bibfnamefont {Q.}~\bibnamefont {{Liu}}}, \ and\ \bibinfo {author}
  {\bibfnamefont {S.}~\bibnamefont {{Wu}}},\ }\href {\doibase
  10.1103/PhysRevA.84.052111} {\bibfield  {journal} {\bibinfo  {journal} {Phys.
  Rev. A}\ }\textbf {\bibinfo {volume} {84}},\ \bibinfo {eid} {052111}
  (\bibinfo {year} {2011})},\ \Eprint {http://arxiv.org/abs/1108.1608}
  {arXiv:1108.1608 [quant-ph]} \BibitemShut {NoStop}%
\bibitem [{\citenamefont {{Nakamura}}\ \emph {et~al.}(2012)\citenamefont
  {{Nakamura}}, \citenamefont {{Nishizawa}},\ and\ \citenamefont
  {{Fujimoto}}}]{Nakamura:2012}%
  \BibitemOpen
  \bibfield  {author} {\bibinfo {author} {\bibfnamefont {K.}~\bibnamefont
  {{Nakamura}}}, \bibinfo {author} {\bibfnamefont {A.}~\bibnamefont
  {{Nishizawa}}}, \ and\ \bibinfo {author} {\bibfnamefont {M.-K.}\ \bibnamefont
  {{Fujimoto}}},\ }\href {\doibase 10.1103/PhysRevA.85.012113} {\bibfield
  {journal} {\bibinfo  {journal} {Phys. Rev. A}\ }\textbf {\bibinfo {volume}
  {85}},\ \bibinfo {eid} {012113} (\bibinfo {year} {2012})},\ \Eprint
  {http://arxiv.org/abs/1108.2114} {arXiv:1108.2114 [quant-ph]} \BibitemShut
  {NoStop}%
\bibitem [{\citenamefont {{Koike}}\ and\ \citenamefont
  {{Tanaka}}(2011)}]{Koike:2011}%
  \BibitemOpen
  \bibfield  {author} {\bibinfo {author} {\bibfnamefont {T.}~\bibnamefont
  {{Koike}}}\ and\ \bibinfo {author} {\bibfnamefont {S.}~\bibnamefont
  {{Tanaka}}},\ }\href {\doibase 10.1103/PhysRevA.84.062106} {\bibfield
  {journal} {\bibinfo  {journal} {Phys. Rev. A}\ }\textbf {\bibinfo {volume}
  {84}},\ \bibinfo {eid} {062106} (\bibinfo {year} {2011})},\ \Eprint
  {http://arxiv.org/abs/1108.2050} {arXiv:1108.2050 [quant-ph]} \BibitemShut
  {NoStop}%
\bibitem [{\citenamefont {{Parks}}\ and\ \citenamefont
  {{Gray}}(2011)}]{Parks:2011}%
  \BibitemOpen
  \bibfield  {author} {\bibinfo {author} {\bibfnamefont {A.~D.}\ \bibnamefont
  {{Parks}}}\ and\ \bibinfo {author} {\bibfnamefont {J.~E.}\ \bibnamefont
  {{Gray}}},\ }\href {\doibase 10.1103/PhysRevA.84.012116} {\bibfield
  {journal} {\bibinfo  {journal} {Phys. Rev. A}\ }\textbf {\bibinfo {volume}
  {84}},\ \bibinfo {eid} {012116} (\bibinfo {year} {2011})},\ \Eprint
  {http://arxiv.org/abs/1104.3639} {arXiv:1104.3639 [quant-ph]} \BibitemShut
  {NoStop}%
\bibitem [{\citenamefont {{Susa}}\ \emph {et~al.}(2012)\citenamefont {{Susa}},
  \citenamefont {{Shikano}},\ and\ \citenamefont {{Hosoya}}}]{Susa:2012}%
  \BibitemOpen
  \bibfield  {author} {\bibinfo {author} {\bibfnamefont {Y.}~\bibnamefont
  {{Susa}}}, \bibinfo {author} {\bibfnamefont {Y.}~\bibnamefont {{Shikano}}}, \
  and\ \bibinfo {author} {\bibfnamefont {A.}~\bibnamefont {{Hosoya}}},\ }\href
  {\doibase 10.1103/PhysRevA.85.052110} {\bibfield  {journal} {\bibinfo
  {journal} {Phys. Rev. A}\ }\textbf {\bibinfo {volume} {85}},\ \bibinfo {eid}
  {052110} (\bibinfo {year} {2012})},\ \Eprint {http://arxiv.org/abs/1203.0827}
  {arXiv:1203.0827 [quant-ph]} \BibitemShut {NoStop}%
\bibitem [{\citenamefont {{Nishizawa}}\ \emph {et~al.}(2012)\citenamefont
  {{Nishizawa}}, \citenamefont {{Nakamura}},\ and\ \citenamefont
  {{Fujimoto}}}]{Nishizawa:2012weak}%
  \BibitemOpen
  \bibfield  {author} {\bibinfo {author} {\bibfnamefont {A.}~\bibnamefont
  {{Nishizawa}}}, \bibinfo {author} {\bibfnamefont {K.}~\bibnamefont
  {{Nakamura}}}, \ and\ \bibinfo {author} {\bibfnamefont {M.-K.}\ \bibnamefont
  {{Fujimoto}}},\ }\href {\doibase 10.1103/PhysRevA.85.062108} {\bibfield
  {journal} {\bibinfo  {journal} {Phys. Rev. A}\ }\textbf {\bibinfo {volume}
  {85}},\ \bibinfo {eid} {062108} (\bibinfo {year} {2012})},\ \Eprint
  {http://arxiv.org/abs/1201.6039} {arXiv:1201.6039 [quant-ph]} \BibitemShut
  {NoStop}%
\bibitem [{\citenamefont {{Tanaka}}\ and\ \citenamefont
  {{Yamamoto}}(2013)}]{Tanaka:2013PRA}%
  \BibitemOpen
  \bibfield  {author} {\bibinfo {author} {\bibfnamefont {S.}~\bibnamefont
  {{Tanaka}}}\ and\ \bibinfo {author} {\bibfnamefont {N.}~\bibnamefont
  {{Yamamoto}}},\ }\href {\doibase 10.1103/PhysRevA.88.042116} {\bibfield
  {journal} {\bibinfo  {journal} {Phys. Rev. A}\ }\textbf {\bibinfo {volume}
  {88}},\ \bibinfo {eid} {042116} (\bibinfo {year} {2013})},\ \Eprint
  {http://arxiv.org/abs/1306.2409} {arXiv:1306.2409 [quant-ph]} \BibitemShut
  {NoStop}%
\bibitem [{\citenamefont {{Ferrie}}\ and\ \citenamefont
  {{Combes}}(2014{\natexlab{a}})}]{Ferrie:2014PRL}%
  \BibitemOpen
  \bibfield  {author} {\bibinfo {author} {\bibfnamefont {C.}~\bibnamefont
  {{Ferrie}}}\ and\ \bibinfo {author} {\bibfnamefont {J.}~\bibnamefont
  {{Combes}}},\ }\href {\doibase 10.1103/PhysRevLett.112.040406} {\bibfield
  {journal} {\bibinfo  {journal} {Physical Review Letters}\ }\textbf {\bibinfo
  {volume} {112}},\ \bibinfo {eid} {040406} (\bibinfo {year}
  {2014}{\natexlab{a}})},\ \Eprint {http://arxiv.org/abs/1307.4016}
  {arXiv:1307.4016 [quant-ph]} \BibitemShut {NoStop}%
\bibitem [{\citenamefont {{Combes}}\ \emph {et~al.}(2014)\citenamefont
  {{Combes}}, \citenamefont {{Ferrie}}, \citenamefont {{Jiang}},\ and\
  \citenamefont {{Caves}}}]{Combes:2014PRA}%
  \BibitemOpen
  \bibfield  {author} {\bibinfo {author} {\bibfnamefont {J.}~\bibnamefont
  {{Combes}}}, \bibinfo {author} {\bibfnamefont {C.}~\bibnamefont {{Ferrie}}},
  \bibinfo {author} {\bibfnamefont {Z.}~\bibnamefont {{Jiang}}}, \ and\
  \bibinfo {author} {\bibfnamefont {C.~M.}\ \bibnamefont {{Caves}}},\ }\href
  {\doibase 10.1103/PhysRevA.89.052117} {\bibfield  {journal} {\bibinfo
  {journal} {Phys. Rev. A}\ }\textbf {\bibinfo {volume} {89}},\ \bibinfo {eid}
  {052117} (\bibinfo {year} {2014})},\ \Eprint {http://arxiv.org/abs/1309.6620}
  {arXiv:1309.6620 [quant-ph]} \BibitemShut {NoStop}%
\bibitem [{\citenamefont {{Vaidman}}(2014)}]{Vaidman:2014comment}%
  \BibitemOpen
  \bibfield  {author} {\bibinfo {author} {\bibfnamefont {L.}~\bibnamefont
  {{Vaidman}}},\ }\href@noop {} {\bibfield  {journal} {\bibinfo  {journal}
  {ArXiv e-prints}\ } (\bibinfo {year} {2014})},\ \Eprint
  {http://arxiv.org/abs/1402.0199} {arXiv:1402.0199 [quant-ph]} \BibitemShut
  {NoStop}%
\bibitem [{\citenamefont {{Kedem}}(2014)}]{Kedem:2014comment}%
  \BibitemOpen
  \bibfield  {author} {\bibinfo {author} {\bibfnamefont {Y.}~\bibnamefont
  {{Kedem}}},\ }\href@noop {} {\bibfield  {journal} {\bibinfo  {journal} {ArXiv
  e-prints}\ } (\bibinfo {year} {2014})},\ \Eprint
  {http://arxiv.org/abs/1402.1352} {arXiv:1402.1352 [quant-ph]} \BibitemShut
  {NoStop}%
\bibitem [{\citenamefont {{Ferrie}}\ and\ \citenamefont
  {{Combes}}(2014{\natexlab{b}})}]{Ferrie:2014comment}%
  \BibitemOpen
  \bibfield  {author} {\bibinfo {author} {\bibfnamefont {C.}~\bibnamefont
  {{Ferrie}}}\ and\ \bibinfo {author} {\bibfnamefont {J.}~\bibnamefont
  {{Combes}}},\ }\href@noop {} {\bibfield  {journal} {\bibinfo  {journal}
  {ArXiv e-prints}\ } (\bibinfo {year} {2014}{\natexlab{b}})},\ \Eprint
  {http://arxiv.org/abs/1402.2954} {arXiv:1402.2954 [quant-ph]} \BibitemShut
  {NoStop}%
\bibitem [{\citenamefont {{Viza}}\ \emph {et~al.}(2014)\citenamefont {{Viza}},
  \citenamefont {{Mart{\'{\i}}nez-Rinc{\'o}n}}, \citenamefont {{Alves}},
  \citenamefont {{Jordan}},\ and\ \citenamefont {{Howell}}}]{Viza:2014}%
  \BibitemOpen
  \bibfield  {author} {\bibinfo {author} {\bibfnamefont {G.~I.}\ \bibnamefont
  {{Viza}}}, \bibinfo {author} {\bibfnamefont {J.}~\bibnamefont
  {{Mart{\'{\i}}nez-Rinc{\'o}n}}}, \bibinfo {author} {\bibfnamefont {G.~B.}\
  \bibnamefont {{Alves}}}, \bibinfo {author} {\bibfnamefont {A.~N.}\
  \bibnamefont {{Jordan}}}, \ and\ \bibinfo {author} {\bibfnamefont {J.~C.}\
  \bibnamefont {{Howell}}},\ }\href@noop {} {\bibfield  {journal} {\bibinfo
  {journal} {ArXiv e-prints}\ } (\bibinfo {year} {2014})},\ \Eprint
  {http://arxiv.org/abs/1410.8461} {arXiv:1410.8461 [quant-ph]} \BibitemShut
  {NoStop}%
\bibitem [{\citenamefont {Braginsky}\ and\ \citenamefont
  {Vorontsov}(1975)}]{Braginsky:1975SvPhU}%
  \BibitemOpen
  \bibfield  {author} {\bibinfo {author} {\bibfnamefont {V.~B.}\ \bibnamefont
  {Braginsky}}\ and\ \bibinfo {author} {\bibfnamefont {Y.~I.}\ \bibnamefont
  {Vorontsov}},\ }\href {\doibase 10.1070/PU1975v017n05ABEH004362} {\bibfield
  {journal} {\bibinfo  {journal} {Soviet Physics Uspekhi}\ }\textbf {\bibinfo
  {volume} {17}},\ \bibinfo {pages} {644} (\bibinfo {year} {1975})}\BibitemShut
  {NoStop}%
\bibitem [{\citenamefont {Caves}\ \emph {et~al.}(1980)\citenamefont {Caves},
  \citenamefont {Thorne}, \citenamefont {Drever}, \citenamefont {Sandberg},\
  and\ \citenamefont {Zimmermann}}]{Caves:1980rv}%
  \BibitemOpen
  \bibfield  {author} {\bibinfo {author} {\bibfnamefont {C.}~\bibnamefont
  {Caves}}, \bibinfo {author} {\bibfnamefont {K.}~\bibnamefont {Thorne}},
  \bibinfo {author} {\bibfnamefont {R.}~\bibnamefont {Drever}}, \bibinfo
  {author} {\bibfnamefont {V.}~\bibnamefont {Sandberg}}, \ and\ \bibinfo
  {author} {\bibfnamefont {M.}~\bibnamefont {Zimmermann}},\ }\href {\doibase
  10.1103/RevModPhys.52.341} {\bibfield  {journal} {\bibinfo  {journal}
  {Rev.Mod.Phys.}\ }\textbf {\bibinfo {volume} {52}},\ \bibinfo {pages} {341}
  (\bibinfo {year} {1980})}\BibitemShut {NoStop}%
%%CITATION = RMPHA,52,341;%%
\bibitem [{\citenamefont {{Aspelmeyer}}\ \emph {et~al.}(2014)\citenamefont
  {{Aspelmeyer}}, \citenamefont {{Kippenberg}},\ and\ \citenamefont
  {{Marquardt}}}]{Aspelmeyer:2014rev}%
  \BibitemOpen
  \bibfield  {author} {\bibinfo {author} {\bibfnamefont {M.}~\bibnamefont
  {{Aspelmeyer}}}, \bibinfo {author} {\bibfnamefont {T.~J.}\ \bibnamefont
  {{Kippenberg}}}, \ and\ \bibinfo {author} {\bibfnamefont {F.}~\bibnamefont
  {{Marquardt}}},\ }\href {\doibase 10.1103/RevModPhys.86.1391} {\bibfield
  {journal} {\bibinfo  {journal} {Reviews of Modern Physics}\ }\textbf
  {\bibinfo {volume} {86}},\ \bibinfo {pages} {1391} (\bibinfo {year}
  {2014})},\ \Eprint {http://arxiv.org/abs/1303.0733} {arXiv:1303.0733
  [cond-mat.mes-hall]} \BibitemShut {NoStop}%
\bibitem [{\citenamefont {{Poot}}\ and\ \citenamefont {{van der
  Zant}}(2012)}]{Poot:2012rev}%
  \BibitemOpen
  \bibfield  {author} {\bibinfo {author} {\bibfnamefont {M.}~\bibnamefont
  {{Poot}}}\ and\ \bibinfo {author} {\bibfnamefont {H.~S.~J.}\ \bibnamefont
  {{van der Zant}}},\ }\href {\doibase 10.1016/j.physrep.2011.12.004}
  {\bibfield  {journal} {\bibinfo  {journal} {Phys. Rep.}\ }\textbf {\bibinfo
  {volume} {511}},\ \bibinfo {pages} {273} (\bibinfo {year} {2012})},\ \Eprint
  {http://arxiv.org/abs/1106.2060} {arXiv:1106.2060 [cond-mat.mes-hall]}
  \BibitemShut {NoStop}%
\bibitem [{\citenamefont {Kimble}\ \emph {et~al.}(2002)\citenamefont {Kimble},
  \citenamefont {Levin}, \citenamefont {Matsko}, \citenamefont {Thorne},\ and\
  \citenamefont {Vyatchanin}}]{Kimble:2000gu}%
  \BibitemOpen
  \bibfield  {author} {\bibinfo {author} {\bibfnamefont {H.}~\bibnamefont
  {Kimble}}, \bibinfo {author} {\bibfnamefont {Y.}~\bibnamefont {Levin}},
  \bibinfo {author} {\bibfnamefont {A.~B.}\ \bibnamefont {Matsko}}, \bibinfo
  {author} {\bibfnamefont {K.~S.}\ \bibnamefont {Thorne}}, \ and\ \bibinfo
  {author} {\bibfnamefont {S.~P.}\ \bibnamefont {Vyatchanin}},\ }\href
  {\doibase 10.1103/PhysRevD.65.022002} {\bibfield  {journal} {\bibinfo
  {journal} {Phys.Rev.}\ }\textbf {\bibinfo {volume} {D 65}},\ \bibinfo {pages}
  {022002} (\bibinfo {year} {2002})},\ \Eprint
  {http://arxiv.org/abs/gr-qc/0008026} {arXiv:gr-qc/0008026 [gr-qc]}
  \BibitemShut {NoStop}%
%%CITATION = GR-QC/0008026;%%
\bibitem [{\citenamefont {{Danilishin}}\ and\ \citenamefont
  {{Khalili}}(2012)}]{Danilishin:2012rev}%
  \BibitemOpen
  \bibfield  {author} {\bibinfo {author} {\bibfnamefont {S.~L.}\ \bibnamefont
  {{Danilishin}}}\ and\ \bibinfo {author} {\bibfnamefont {F.~Y.}\ \bibnamefont
  {{Khalili}}},\ }\href {\doibase 10.12942/lrr-2012-5} {\bibfield  {journal}
  {\bibinfo  {journal} {Living Reviews in Relativity}\ }\textbf {\bibinfo
  {volume} {15}},\ \bibinfo {pages} {5} (\bibinfo {year} {2012})},\ \Eprint
  {http://arxiv.org/abs/1203.1706} {arXiv:1203.1706 [quant-ph]} \BibitemShut
  {NoStop}%
\bibitem [{\citenamefont {{Braginsky}}\ \emph {et~al.}(1992)\citenamefont
  {{Braginsky}}, \citenamefont {{Khalili}},\ and\ \citenamefont
  {{Thorne}}}]{Braginsky:book}%
  \BibitemOpen
  \bibfield  {author} {\bibinfo {author} {\bibfnamefont {V.~B.}\ \bibnamefont
  {{Braginsky}}}, \bibinfo {author} {\bibfnamefont {F.~Y.}\ \bibnamefont
  {{Khalili}}}, \ and\ \bibinfo {author} {\bibfnamefont {K.~S.}\ \bibnamefont
  {{Thorne}}},\ }\href@noop {} {\emph {\bibinfo {title} {Quantum Measurement,
  by Vladimir B.~Braginsky , Farid Ya Khalili , Kip S.~Thorne, Cambridge, UK:
  Cambridge University Press, 1992}}}\ (\bibinfo {year} {1992})\BibitemShut
  {NoStop}%
\bibitem [{\citenamefont {{Giovannetti}}\ \emph {et~al.}(2004)\citenamefont
  {{Giovannetti}}, \citenamefont {{Lloyd}},\ and\ \citenamefont
  {{Maccone}}}]{Giovannetti:2004rev}%
  \BibitemOpen
  \bibfield  {author} {\bibinfo {author} {\bibfnamefont {V.}~\bibnamefont
  {{Giovannetti}}}, \bibinfo {author} {\bibfnamefont {S.}~\bibnamefont
  {{Lloyd}}}, \ and\ \bibinfo {author} {\bibfnamefont {L.}~\bibnamefont
  {{Maccone}}},\ }\href {\doibase 10.1126/science.1104149} {\bibfield
  {journal} {\bibinfo  {journal} {Science}\ }\textbf {\bibinfo {volume}
  {306}},\ \bibinfo {pages} {1330} (\bibinfo {year} {2004})},\ \Eprint
  {http://arxiv.org/abs/quant-ph/0412078} {quant-ph/0412078} \BibitemShut
  {NoStop}%
\bibitem [{\citenamefont {{Pang}}\ \emph {et~al.}(2014)\citenamefont {{Pang}},
  \citenamefont {{Dressel}},\ and\ \citenamefont {{Brun}}}]{Pang2014a}%
  \BibitemOpen
  \bibfield  {author} {\bibinfo {author} {\bibfnamefont {S.}~\bibnamefont
  {{Pang}}}, \bibinfo {author} {\bibfnamefont {J.}~\bibnamefont {{Dressel}}}, \
  and\ \bibinfo {author} {\bibfnamefont {T.~A.}\ \bibnamefont {{Brun}}},\
  }\href {\doibase 10.1103/PhysRevLett.113.030401} {\bibfield  {journal}
  {\bibinfo  {journal} {Physical Review Letters}\ }\textbf {\bibinfo {volume}
  {113}},\ \bibinfo {eid} {030401} (\bibinfo {year} {2014})},\ \Eprint
  {http://arxiv.org/abs/1401.5887} {arXiv:1401.5887 [quant-ph]} \BibitemShut
  {NoStop}%
\bibitem [{\citenamefont {{Pang}}\ and\ \citenamefont
  {{Brun}}(2014)}]{Pang2014b}%
  \BibitemOpen
  \bibfield  {author} {\bibinfo {author} {\bibfnamefont {S.}~\bibnamefont
  {{Pang}}}\ and\ \bibinfo {author} {\bibfnamefont {T.~A.}\ \bibnamefont
  {{Brun}}},\ }\href@noop {} {\bibfield  {journal} {\bibinfo  {journal} {ArXiv
  e-prints}\ } (\bibinfo {year} {2014})},\ \Eprint
  {http://arxiv.org/abs/1409.2567} {arXiv:1409.2567 [quant-ph]} \BibitemShut
  {NoStop}%
\bibitem [{\citenamefont {{Zhang}}\ \emph {et~al.}(2013)\citenamefont
  {{Zhang}}, \citenamefont {{Datta}},\ and\ \citenamefont
  {{Walmsley}}}]{Zhang2013arXiv}%
  \BibitemOpen
  \bibfield  {author} {\bibinfo {author} {\bibfnamefont {L.}~\bibnamefont
  {{Zhang}}}, \bibinfo {author} {\bibfnamefont {A.}~\bibnamefont {{Datta}}}, \
  and\ \bibinfo {author} {\bibfnamefont {I.~A.}\ \bibnamefont {{Walmsley}}},\
  }\href@noop {} {\bibfield  {journal} {\bibinfo  {journal} {ArXiv e-prints}\ }
  (\bibinfo {year} {2013})},\ \Eprint {http://arxiv.org/abs/1310.5302}
  {arXiv:1310.5302 [quant-ph]} \BibitemShut {NoStop}%
\bibitem [{\citenamefont {{Jordan}}\ \emph
  {et~al.}(2014{\natexlab{b}})\citenamefont {{Jordan}}, \citenamefont
  {{Tollaksen}}, \citenamefont {{Troupe}}, \citenamefont {{Dressel}},\ and\
  \citenamefont {{Aharonov}}}]{Jordan2014QSMF}%
  \BibitemOpen
  \bibfield  {author} {\bibinfo {author} {\bibfnamefont {A.~N.}\ \bibnamefont
  {{Jordan}}}, \bibinfo {author} {\bibfnamefont {J.}~\bibnamefont
  {{Tollaksen}}}, \bibinfo {author} {\bibfnamefont {J.~E.}\ \bibnamefont
  {{Troupe}}}, \bibinfo {author} {\bibfnamefont {J.}~\bibnamefont {{Dressel}}},
  \ and\ \bibinfo {author} {\bibfnamefont {Y.}~\bibnamefont {{Aharonov}}},\
  }\href@noop {} {\bibfield  {journal} {\bibinfo  {journal} {ArXiv e-prints}\ }
  (\bibinfo {year} {2014}{\natexlab{b}})},\ \Eprint
  {http://arxiv.org/abs/1409.3488} {arXiv:1409.3488 [quant-ph]} \BibitemShut
  {NoStop}%
\end{thebibliography}%

\end{document}